\documentclass[twocolumn]{openjournal}
\usepackage{lipsum}

\usepackage{xcolor}
\usepackage{textgreek}
\usepackage[utf8]{inputenc}
\usepackage[english]{babel}

\usepackage{hyperref}
\hypersetup{
    unicode, 
    colorlinks=true,
    linkcolor=linkcolor,
    citecolor=linkcolor,
    filecolor=linkcolor,
    urlcolor=linkcolor,
}
\usepackage{color,colortbl}
\definecolor{linkcolor}{rgb}{0.0,0.3,0.5}
\usepackage{tensind}
\tensordelimiter{?}
\DeclareGraphicsExtensions{.bmp,.png,.jpg,.pdf}
\usepackage{verbatim}
\usepackage[normalem]{ulem}
\usepackage{orcidlink}
\usepackage{soul}

\graphicspath{ {./figs/} }

\newcommand{\lla}{\left\langle}
\newcommand{\rra}{\right\rangle}
\newcommand{\be}{\begin{equation}}
\newcommand{\ee}{\end{equation}}
\newcommand{\bea}{\begin{eqnarray}}
\newcommand{\eea}{\end{eqnarray}}
\newcommand{\dv}[2]{\frac{d#1}{d#2}}
\newcommand{\pdv}[2]{\frac{\partial #1}{\partial #2}}
\newcommand{\bnabla}{\mathbf{\nabla}}
\newcommand{\bfv}{\mathbf{v}}
\newcommand{\br}{\mathbf{r}}
\newcommand{\bJ}{\mathbf{J}}
\newcommand{\bB}{\mathbf{B}}
\newcommand{\bF}{\mathbf{F}}

\begin{document}

\title{Non-linear Dynamical Stability of Magnetic Polytropes}

\author{Bryan M. Johnson\orcidlink{0000-0001-8973-0680}}
\email{johnsonbryanmark@gmail.com}
\affiliation{University of Alabama in Huntsville}

\begin{abstract}

This work analyzes the non-linear dynamical stability of ideal-gas polytropes under homologous flow. A non-constant density profile requires the inclusion of magnetic fields, which is done by introducing a mean-field model that treats the spherically-averaged radial Lorentz force self-consistently and has the following properties: 1) The only essential simplifications are the Cowling approximation and a dominant radial flow. 2) The average radial Lorentz force due to an isotropic field is $-\frac13 dP_B/dr$, not $-dP_B/dr$ as is typically assumed. 3) A central peak in the magnetic field requires isotropy there; all other configurations are zero at the origin due to magnetic tension. 4) Solutions with negligible surface fields require $\gtrsim1/2$ of the magnetic energy to be in the radial component. 5) Solutions that resemble Lane-Emden solutions are restricted to $\gamma = 4/3$, where $\gamma$ is the material adiabatic index, and exhibit either collapse or escape. 6) Solutions for general $\gamma$ have a harmonic enthalpy profile and allow for non-linear radial pulsations. 7) A harmonic-enthalpy homologous flow becomes unbound when an overpressure satisfies $\delta = \Delta P_0/P_{\rm eq} > \frac{3\gamma - 4}{1 + 3(\gamma-1)\alpha_0}$, where $P$ is the total pressure, $P_{\rm eq}$ is its equilibrium value, $\alpha$ is the ratio of radiation to material pressure, and a zero subscript denotes minimum volume. This indicates that radiation pressure can unbind a linearly-stable polytrope in the presence of small but finite radial perturbations. The condition to unbind a fully-ionized $n = 3$ polytrope with $2/3$ of its magnetic energy in the radial component is $\delta \gtrsim 0.15\mu^{-1}\left(300M_\odot/M\right)^{1/2}$, where $\mu$ is the mean molecular weight. This non-linear dynamical instability threshold may have some relevance for mass loss in and dispersal of evolved high-mass stars.

\end{abstract}

% Write your keywords here

\maketitle

\section{Introduction}
\label{sec:intro}

Stellar equilibria with an ideal gas equation of state are dynamically stable to infinitesimal (linear amplitude) perturbations when $\gamma > 4/3$, where $\gamma$ is the material adiabatic index \citep{Chandra1939}. The ideal-gas assumption remains a useful guide for non-ideal gases, as various physical effects can be viewed as reducing the effective adiabatic index below $4/3$, resulting in instability \citep{Chandra1964,Baranov2013}. Where the assumption of linearity breaks down is less clear, as determining a non-linear stability boundary generally requires a complicated non-linear analysis or direct numerical simulation \citep{Rein2003,Luo2008}, and the parameter space is vast. Since evolved high-mass stars are often near their linear stability boundary, understanding the location of non-linear stability boundaries could impact our understanding of their dynamical behavior, including mass loss, explosions, and the formation of compact remnants. While large departures from equilibrium are unlikely, perturbations that are small, but not infinitesimally so, may be relevant in these systems if their evolution brings them near a non-linear rather than a linear stability boundary. Understanding the stability of these systems in that case would require quantifying the amplitude of any potential perturbations.

To produce a tractable problem, this work will investigate the non-linear dynamical stability of the following idealized configuration: homologous flow of an ideal-gas polytrope with a spherically-averaged magnetic field. Although the polytropic assumption is not strictly necessary, it allows most of the results to be obtained analytically. In his review of pulsating stars, \cite{Cox1974} derives solutions for non-linear radial pulsations similar to the ones derived here, but in the absence of magnetic fields these solutions are isochoric. The solutions described below include magnetic fields, and as a result have density profiles that, while still idealized, are more realistic than an isochoric profile. At the same time, while homologous-flow solutions that resemble Lane-Emden solutions can be obtained, they require $\gamma = 4/3$. Assessing non-linear dynamical stability for general $\gamma$ requires an enthalpy profile that is spatially harmonic and results in a force balance that differs from the standard idealized form.

Since the magnetic field model used here has aspects that are new, care will be taken to describe it thoroughly. Due to their complexity and inaccessibility to direct observation, internal magnetic fields are difficult to incorporate into models of stellar structure. In general, they require a three-dimensional treatment, and disorder seems to be an essential characteristic \citep{Parker1957}. Simplified magnetic configurations are unstable \citep{Tayler1973,Wright1973,MT1973,Braithwaite2006}, and the stable numerical equilibria that have been found consist of twisted axisymmetric flux loops or randomly-oriented flux loops that meander throughout the star \citep{BS2004,BN2006,BS2006,Mitchell2015,BS2017,Becerra2022a,Becerra2022b}. Most stellar magnetic fields are thought to arise from the action of a dynamo, which adds the additional complexity of turbulent velocity fluctuations \citep{Charbonneau2014}. Despite impressive advances in numerical techniques and computational resources, however, dynamic range limitations will remain a barrier to numerical magnetic models of stellar structure for the foreseeable future \citep{Browning2023}.

Simplified models of internal stellar magnetic fields sacrifice rigor for tractability, and in many cases are the only viable alternative to ignoring magnetic fields altogether. The reduced-order model for internal stellar magnetic fields developed by \cite{LS1995} treats the magnetic field somewhat heuristically as a thermodynamic variable, and has been used to assess the anamolously large radii of some low mass stars \citep{FC2012,FC2013,FC2014,MM2001,MM2017}. \cite{Gupta2020} and \cite{Karinkuzhi2024} use an analytical expression for the magnetic field as a function of density in their one-dimensional models for the mass-radius relation of magnetized white dwarfs, based on the impact of strong magnetic fields on the equation of state of degenerate matter \citep{Bandyopadhyay1997,Das2012,Das2014}. In their analysis of the impact of internal magnetic fields on the depressed dipole modes seen in red giant stars, \cite{Fuller2015} and \cite{Cantiello2016} introduce a core magnetic field and provide constraints on its amplitude, but leave the internal structure otherwise unspecified. Analytical models for internal magnetic fields assume axisymmetry and have limited utility due to their instability \citep{BS2017}, although some stable analytical equilibria have been found \citep{Lyutikov2010,DM2010,DBM2010,Akgun2013}.

All of the simplified models described above have sacrificed more rigor than is necessary, however, since none of them self-consistently apply the spherically-averaged radial equation of motion. This work will derive that equation as the basis for a self-consistent mean-field model of internal stellar magnetic fields. While this approach is in one sense more restrictive than the assumption of axisymmetry because it reduces the equations to one dimension, it is less restrictive in the sense that the average angular variables are retained exactly in the radial equation. Angular information is lost, and ans$\ddot{\rm a}$tze for the average variables must be introduced in order to produce a complete model, but these model closures can be directly compared to numerical experiments to assess their validity. The applicability of the approach to non-axisymmetric equilibria is a significant advantage over existing analytical approaches \citep{Becerra2022b}.

The only necessary simplifications to make in order to derive a self-consistent model are 1) apply the Cowling approximation, and 2) assume a dominant mean flow. The present analysis ignores velocity fluctuations altogether, which limits its application to non-convective stars, or at least to stars in which convective velocities are subsonic and sub-Alfv\'enic. The rigorous form for spherically-averaged incompressive velocity fluctuations is included for reference. The basic equations and assumptions are outlined in \S\ref{sec:equations}, results are given in \S\ref{sec:results}, and a discussion and directions for future work is contained in \S\ref{sec:discussion}.

\section{Equations and assumptions}\label{sec:equations}

The equations of ideal magnetohydrodynamics for a self-gravitating, adiabatic ideal gas in Lorentz-Heaviside units are:
\be\label{mass}
\dv{\rho}{t} = -\rho \bnabla \cdot \bfv,
\ee
\be\label{momentum}
\rho\dv{\bfv}{t} = -\bnabla P - \rho \bnabla \Phi + \bJ \times \bB,
\ee
\be\label{induction}
\dv{\bB}{t} = -\bB \bnabla \cdot  \bfv +  \bB \cdot \bnabla \bfv,
\ee
\be\label{poisson}
\nabla^2 \Phi = 4\pi G \rho,
\ee
\be\label{adiabatic}
\dv{s}{t} = 0,
\ee
where $\rho$ is the mass density, $\bfv$ is the velocity, $P$ is the pressure, $\Phi$ is the gravitational potential, $\bB$ is the magnetic field, $\bJ = \bnabla \times \bB$ is the current density, $s$ is the specific entropy, and $d/dt$ is the Lagrangian derivative following a fluid element. These equations are supplemented by the divergence-free constraint on the magnetic field.

\cite{Woltjer1962} has demonstrated that in the absence of strong surface currents, the outer equipotential surfaces of stellar equilibria will be spherical. This motivates the development and use of a mean-field model in a spherical coordinate system, rather than the usual decomposition into poloidal and toroidal components. The spherically-averaged radial equation of motion is
\be\label{momentumr}
\lla\rho\dv{\bfv}{t}\rra_r = -\pdv{\lla P \rra}{r} - \lla\rho \pdv{\Phi}{r}\rra + \lla\bJ \times \bB\rra_r,
\ee
where
\[
\lla f \rra \equiv \frac{1}{4\pi} \int f \, d\Omega = \int_0^{2\pi} d\phi \int_0^\pi f\sin \theta \, d\theta.
\]
Applying the Cowling approximation and assuming $\bfv = \lla v\rra \hat{\mathbf{r}}$, equation (\ref{momentumr}) reduces to
\be\label{momentumr2}
\lla\rho\rra \dv{\lla v\rra}{\lla t\rra} = -\pdv{\lla P \rra}{r} - \lla\rho\rra\pdv{\lla\Phi\rra}{r} + \lla\bJ \times \bB\rra_r,
\ee
where
\[
\dv{}{\lla t\rra} \equiv \pdv{}{t} + \lla v\rra \pdv{}{r}
\]
is the derivative following the mean flow. It is shown in Appendix~\ref{sec:angular_average} that the spherically-averaged radial Lorentz force density is given exactly by
\be\label{eq:LFr1}
\lla  \bJ \times \bB \rra_r = -\frac{1}{r^2}\pdv{}{r}\left(r^2   \lla P_{B\sphericalangle}\rra\right) + \frac{1}{r^4}\pdv{}{r}\left(r^4  \lla P_{Br}\rra\right),
\ee
where $P_{Bi} \equiv B_i^2/2$ is the magnetic pressure associated with magnetic field component $B_i$, and $P_{B\sphericalangle} \equiv P_{B\theta} + P_{B\phi}$
is the angular magnetic pressure. An alternative equivalent formulation is
\be\label{eq:LFr2}
\lla\bJ \times \bB\rra_r = \left(2\chi-1\right)\pdv{\lla P_{B}\rra}{r} + 2\left(\frac{3\chi-1}{r} + \pdv{\chi}{r}\right)\lla P_{B}\rra,
\ee
where
\[
\chi\left(r,t\right) \equiv \frac{\lla P_{Br}\rra}{\lla P_{B}\rra}
\]
is the ratio of the average radial magnetic energy to the average total magnetic energy. Hydrostatic equilibrium implies $\chi = \chi(r)$, and by definition $0 \leq \chi \leq 1$.

Several characteristics of the spherically-averaged Lorentz force can be seen immediately. 1) The only self-consistent model for which magnetic tension can be neglected and the average Lorentz force be reduced to a simple pressure is $\chi = 1/3$, in which case
\[
\lla\bJ \times \bB\rra_r = -\frac{1}{3}\pdv{\lla P_{B}\rra}{r}.
\]
2) For $\chi = 0$, the Lorentz force is
\[
\lla\bJ \times \bB\rra_r = -\pdv{\lla P_{B}\rra}{r} - \frac{2\lla P_B\rra}{r},
\]
and it can be shown that the magnetic tension, which cannot be neglected, results in a Lorentz force that is directed inwards. In fact, the Lorentz force is generally directed inward (outward) for $\chi < 1/2$ ($\chi > 1/2$). 3) Only $\chi = 1/3$ allows for a magnetic pressure profile that peaks at the origin. When $\chi \neq 1/3$, the magnetic pressure must have a minimum at the origin, since $P_B = 0$ is required at $r = 0$ in order for the magnetic tension to be finite there.\footnote{Note that while $\chi = 1/3$ is consistent with an isotropic field, it only implies that $1/3$ of the magnetic energy is in the radial component; the remaining $2/3$ can be distributed arbitrarily between the angular components.}

An expression for $\chi$ can be derived under the assumption of a radial mean flow using equation (\ref{induction}) in the form
\be\label{eq:PBr}
\dv{\lla P_{Br}\rra}{t} = -2\lla P_{Br}\rra\bnabla \cdot \bfv + 2\lla P_{Br}\rra\pdv{v}{r},
\ee
\be\label{eq:PBa}
\dv{\lla  P_{B\sphericalangle}\rra}{t} = -2\lla P_{B\sphericalangle}\rra\bnabla \cdot \bfv + 2\lla P_{Br}\rra\frac{v}{r},
\ee
which equations can be integrated to give the spherically-averaged flux-freezing conditions for a radial flow:
\[
\lla P_{Br}\rra = \lla P_{Br0}\rra\left(\frac{r_0}{r}\right)^4, \;\; \lla P_{B\sphericalangle}\rra = \lla P_{B\sphericalangle 0}\rra \left(\frac{\rho r}{\rho_0 r_0}\right)^2.
\]
These can be combined to give
\be\label{chi}
\chi = \frac{\chi_0}{\chi_0+\left(1-\chi_0\right)\frac{\rho^2 r^6}{\rho_0^2 r_0^6}} = \frac{\chi_0}{\chi_0+\left(1-\chi_0\right)\left(\pdv{\ln r_0}{\ln r}\right)^2}.
\ee
Here and throughout a zero subscript denotes the spatial profile of a quantity at an arbitrary $t = 0$, and can be regarded as a Lagrangian label on a fluid element ($\chi_0$ and $\rho_0$ in the above expressions are fields, not constants). Expression (\ref{chi}) shows that (compression, expansion) drives $\chi$ to ($0, 1$), and $\chi = \chi_0(r_0)$ for homologous flow.

At this point, the bracket notation will be removed, with the understanding that all of the derived results refer to spherically-averaged quantities. Combining equations (\ref{eq:PBr}) and (\ref{eq:PBa}), and rewriting (\ref{chi}) as an evolution equation, yields the final spherically-averaged equations for a radial flow under the Cowling approximation:
\bea\label{momentumrf}
\rho \dv{v}{t} = -\pdv{P}{r} - \rho\pdv{\Phi}{r} + \left(2\chi-1\right)\pdv{P_{B}}{r}\nonumber \\
 + 2\left(\frac{3\chi-1}{r} + \pdv{\chi}{r}\right)P_{B},
\eea
\be\label{chieq}
\dv{\chi}{t} = 2\chi\left(1-\chi\right)\left(\pdv{v}{r}-\frac{v}{r}\right),
\ee
\be\label{PBeq}
\dv{P_{B}}{t} = -2P_{B}\bnabla \cdot \bfv + 2\chi P_{B}\left(\pdv{v}{r} + \frac{v}{r}\right).
\ee
Alternatively, one could use the Lorentz force in the form given by (\ref{eq:LFr1}) in equation (\ref{momentumrf}), along with equations (\ref{eq:PBr}) and (\ref{eq:PBa}). These equations can be used to augment a one-dimensional stellar structure model with magnetic fields in a self-consistent manner. The angular average of equations (\ref{mass}), (\ref{poisson}), and (\ref{adiabatic}) have the same form as the original for a flow dominated by a radial mean flow. 

Homologous flow is expressed mathematically as $r = r_0 a(t)$, where the scale factor $a < 1$ ($a > 1$) corresponds to compression (expansion). The velocity field for spherical homologous flow is $v = r_0 \dot{a}$, where a dot denotes a time derivative. Under this assumption, equations (\ref{mass}),  (\ref{induction}), and (\ref{poisson}) can be integrated to give
\be\label{eq:homologous_scalings}
\rho = \frac{\rho_0(r_0)}{a^3}, \; B = \frac{B_{0}(r_0)}{a^2},\;\Phi = \frac{\Phi_0(r_0)}{a},
\ee
and $\chi$ is independent of time. The hydrostatic limit corresponds to $a = 1$ for all time. 

The pressure is taken to be the sum of material and radiation pressure, $P = P_m + P_r$:
\[
P = P_m\left(1+\alpha\right) = P_0\frac{1+\alpha}{1+\alpha_0}\left(\frac{\alpha}{\alpha_0}\right)^\frac{1}{3}\left(\frac{\rho}{\rho_0}\right)^\frac{4}{3},
\]
where $\alpha \equiv P_r/P_m$, $P_m = (\gamma-1)e_m$, $\gamma$ is the material adiabatic index, $e_m$ is the thermal energy density of the material, $P_r = \frac13 e_r$ is the radiation pressure under the Eddington approximation, and $e_r = a_B T^4$ is the energy density of the radiation \citep{Johnson2009}. The equation of motion for spherical homologous flow under these assumptions is
\bea
\rho_0 r_0\ddot{a} = 
-\frac{1}{a^2}\left(\frac{\alpha}{\alpha_0}\right)^\frac{1}{3}\frac{1+\alpha}{1+\alpha_0}\dv{P_0}{r_0} - \frac{1}{a^2}\left(\rho_0\dv{\Phi_0}{r_0} \right. \nonumber \\ \left.
+ \frac{1}{r_0^2}\dv{}{r_0}\left[r_0^2 P_{B\sphericalangle 0}\right] - \frac{1}{r_0^4}\dv{}{r_0}\left[r_0^4 P_{Br0}\right]\right), \;\;\;\;\;\;\;
\eea
where
\[
\alpha = \frac{1}{12(\gamma-1)}W\left(\frac{12[\gamma-1]\alpha_0}{a^{3[3\gamma-4]}} e^{12[\gamma-1]\alpha_0}\right)
\]
and $W(x)$ is the Lambert-$W$ function \citep{Johnson2009}. For $\alpha_0 = 0$, $P = P_0/a^{3\gamma}$. 

\begin{deluxetable*}{c|c}
\tablecaption{Governing equations for spherically-averaged homologous flow.
\label{tab:Variable_star}}
\setlength{\extrarowheight}{10pt}
\startdata
$\gamma = 4/3$ equations & $\displaystyle a^2\ddot{a} = \delta\omega_d^2 = -\frac{1}{\rho_0 r_0}\left(\dv{P_0}{r_0} + \rho_0\dv{\Phi_0}{r_0} + \frac{1}{r_0^2}\dv{}{r_0}\left[r_0^2 P_{B\sphericalangle 0}\right] - \frac{1}{r_0^4}\dv{}{r_0}\left[r_0^4 P_{Br0}\right]\right)$\\[10pt]
Harmonic enthalpy spatial equations & $\displaystyle
-\frac{1}{\rho_0 r_0(1+\delta)}\dv{P_0}{r_0} = \omega_{d}^2 = \frac{1}{r_0}\dv{\Phi_0}{r_0} + \frac{1}{\rho_0 r_0}\left(\frac{1}{r_0^2}\dv{}{r_0}\left[r_0^2 P_{B\sphericalangle 0}\right] - \frac{1}{r_0^4}\dv{}{r_0}\left[r_0^4  P_{Br0}\right]\right)$ \\[10pt]
Harmonic enthalpy temporal equation & $\displaystyle
\frac{\ddot{a}}{\omega_d^2} = 
\frac{1+\delta}{a^2}\left(\frac{\alpha}{\alpha_0}\right)^\frac{1}{3}\frac{1+\alpha}{1+\alpha_0} - \frac{1}{a^2} \;\; \to \;\; \frac{\ddot{a}}{\omega_d^2} = 
\frac{1+\delta}{a^{3\gamma-2}} - \frac{1}{a^2} \;\; {\rm for} \;\; \alpha_0 = 0$\\[10pt]
Departure from hydrostatic equilibrium & $\displaystyle
\delta = \frac{3(\gamma-1)E_{m0} + E_{r0} + E_{B0} + E_{g0}}{I_0\omega_d^2}$\\[10pt]
\enddata
\end{deluxetable*}

\section{Results}\label{sec:results}

There are two broad classes of solutions to the homologous equation of motion, both of which reduce to hydrostatic solutions in the limit $\ddot{a} = \dot{a} = 0$. The first class is obtained by constraining the time dependence of the force terms to be the same, which requires $\gamma = 4/3$. This also requires introducing a separation constant, defined here as $\delta\omega_d^2$ (with $\omega_d^2 > 0$), where $\omega_d$ is the dynamical frequency, and $\delta$ is measure of the departure from hydrostatic equilibrium which is not required to be small. The resulting temporal and spatial equations are given in the first row of Table~\ref{tab:Variable_star}. There are two  temporal outcomes for gravitational homologous flow under this assumption: collapse to a singularity (for $\delta < 0$) or escape to infinity (for $\delta > 0$) \citep{GW1980,Low1982a,Low1982b,Low1984a,Low1984b,Low1992,Hennebelle2001,Hennebelle2003}. In the absence of magnetic fields, $\omega_d = \omega_g$, where
\[
\omega_{g} \equiv \sqrt{\frac{4\pi G \rho_{c0}}{3}}
\]
is the inverse of the central free-fall time  at $t = 0$. Here and throughout a $c$ subscript denotes a value at the origin.

The second class of solutions is obtained by setting each of the spatial terms on the right-hand side of the equation of motion to a separate constant: $(1+\delta)\omega_d^2$ for the pressure term, and $\omega_d^2$ for the magnetic and gravitational terms (which have the same time dependence). This is effectively choosing a spatial separation constant rather than a temporal one (see the second row of Table~\ref{tab:Variable_star}), and it both constrains the spatial profiles more and allows for more flexibility in the temporal behavior. In particular, it opens up the possibility of a third type of gravitational flow: persistent pulsations. Since the spatial constraint results in an enthalpy profile that is harmonic, these shall be referred to as harmonic entalphy solutions. The temporal equation for this class of solutions is provided in row 3 of Table~\ref{tab:Variable_star}. It can be seen from this equation that an additional constraint to produce a temporal equation that is independent of space is that $\alpha_0$ must be a constant. Using the virial theorem (see Appendix~\ref{sec:virial}), the departure from hydrostatic equilibrium can be expressed as shown in row 4 of Table~\ref{tab:Variable_star}, where each $E$ is a volume-integrated energy component and $I$ is the moment of inertia. For the harmonic-enthalpy solutions, $\delta$ can also be expressed as an overpressure ($\delta = \Delta P_0/P_{\rm eq}$, where $P_{eq}$ is the equlibrium pressure profile). After a discussion of the stability landscape in the next subsection, the detailed time and space dependence of the two classes of solutions will be analyzed in the following subsections. 

\subsection{Stability landscape}\label{sec:stability}

The integral of the harmonic-enthalpy temporal equation in Table~\ref{tab:Variable_star} is
\[
\frac{\dot{a}^2}{2\omega_d^2} = \int\frac{1+\delta}{a^2}\left(\frac{\alpha}{\alpha_0}\right)^\frac{1}{3}\frac{1+\alpha}{1+\alpha_0}\,da + \frac{1}{a}.
\]
Converting the integral over $a$ to an integral over $\alpha$ using
\[
\frac{d\alpha}{\alpha} = -\frac{3(3\gamma-4)}{1+12(\gamma-1)\alpha}\frac{da}{a},
\]
this can be shown to be
\bea\label{eq:temporalHE}
\frac{\dot{a}^2}{2\omega_d^2} =  \frac{1+\delta}{1+\alpha_0}\left(\alpha_0 + \frac{1}{3(\gamma-1)}
\right) - 1
\nonumber \\
+ a^{-1} \left(1 - \frac{1+\delta}{1+\alpha_0}\left[\frac{\alpha}{\alpha_0}\right]^{\frac13}\left[\alpha+\frac{1}{3(\gamma-1)}\right]\right),
\eea
where
\[
a^{-1} = \left(\frac{\alpha}{\alpha_0}e^{12[\gamma-1][\alpha-\alpha_0]}\right)^\frac{1}{3(3\gamma-4)},
\]
a result that is equivalent to the statement of energy conservation $K = E_0 - E$ (with $K_0 = 0$), where $E = E_m + E_r + E_B + E_g$ is the total potential energy (see Appendix~\ref{sec:virial}). The integration constant in equation (\ref{eq:temporalHE}) has been set to zero, which ensures that the flow has at least one turning point. The $\alpha_0 = 0$ limit of equation (\ref{eq:temporalHE}) is
\be\label{eq:temporalHE0}
\frac{\dot{a}^2}{2\omega_d^2} = -\frac{1+\delta}{3(\gamma-1)}\left(\frac{1}{a^{3(\gamma-1)}} - 1\right) + \frac{1}{a} - 1.
\ee

\begin{figure}
\includegraphics[width=\columnwidth]{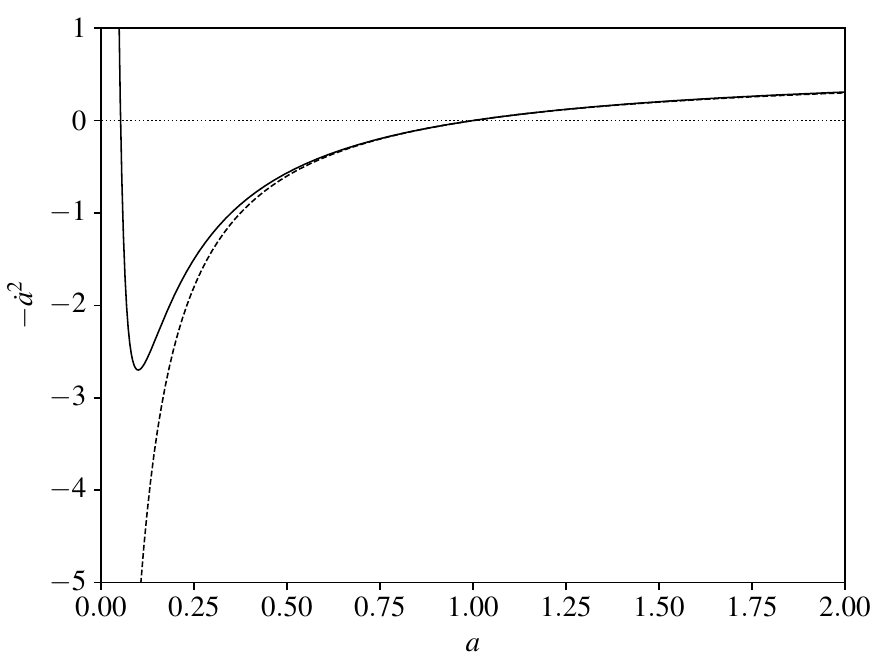}
\caption{\label{fig:TypeIIvelocity}Energy integral with $\delta = -0.9$ and $\gamma = 5/3$ (\emph{solid}) and $4/3$ (\emph{dashed}).}
\end{figure}

\begin{figure}
\includegraphics[width=\columnwidth]{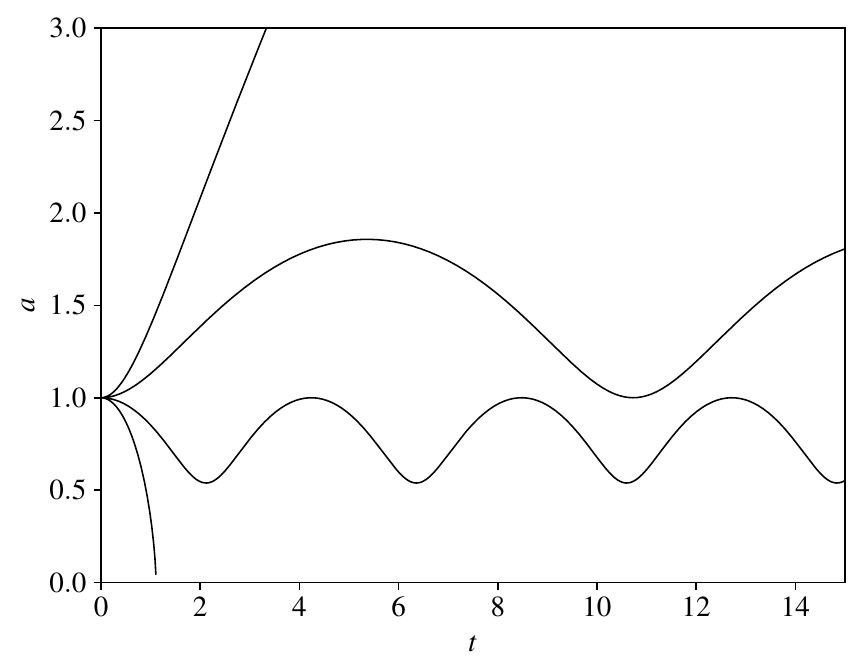}
\caption{\label{fig:Four_solutions}Evolution of the scale factor for $\gamma = 5/3$ and $\delta = -1$, $-0.3$, $0.3$, and $1$ (from bottom to top).}
\end{figure}

Figure~\ref{fig:TypeIIvelocity} shows two examples of this energy integral, demonstrating that for some parameter values there can be two turning points in the flow. This is the signature of a pulsating equilibrium: the flow is trapped indefinitely between a maximum and minimum radius. Figure~\ref{fig:Four_solutions} shows the time dependence of the scale factor for the three types of solutions: collapse, pulsation, and escape. To determine what regions of parameter space contain pulsating solutions, one can analyze the equation of motion and minimum energy integral either graphically or algebraically. A collapse solution will have $\dot{a}^2 > 0$ as $a \rightarrow 0$ and $\ddot{a} < 0$ at $a = 1$. An escape solution will have $\dot{a}^2 > 0$ as $a \rightarrow \infty$ and $\ddot{a} > 0$ at $a = 1$. A pulsating solution that compresses from hydrostatic equilibrium ($0 < a < 1$) will have $\dot{a}^2 > 0$ as $a \rightarrow 0$ and $\ddot{a} > 0$ at $a = 1$, and a pulsating solution that expands from hydrostatic equilibrium ($1 < a < \infty$) will have $\dot{a}^2 < 0$ as $a \rightarrow \infty$ and $\ddot{a} > 0$ at $a = 1$. 

\begin{figure}
\includegraphics[width=\columnwidth]{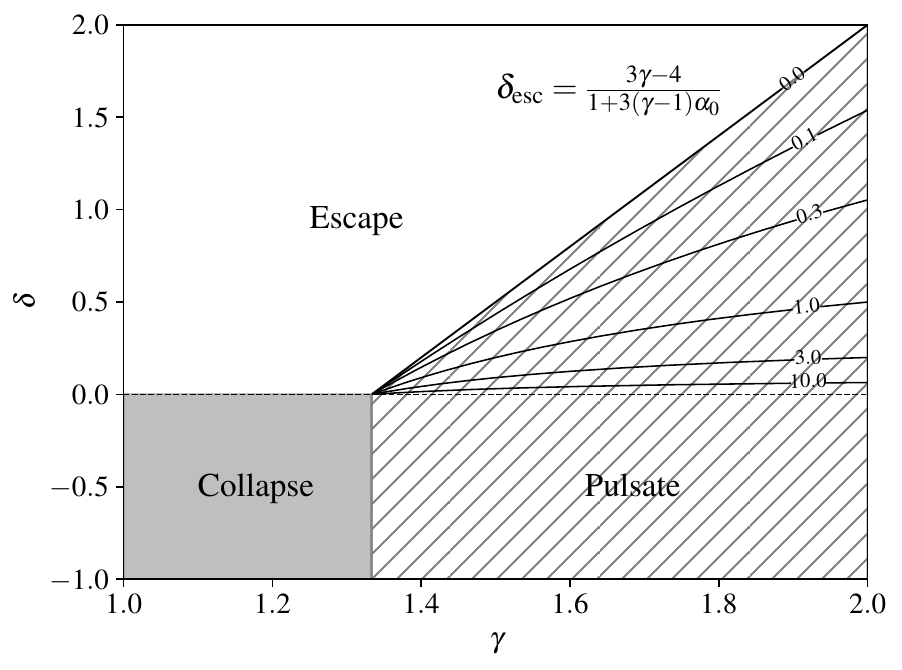}
\caption{\label{fig:TypeIIlandscape3}Landscape of harmonic enthalpy solutions. The dashed line is the locus of hydrostatic solutions, and the solid lines are thresholds for escape labeled by $\alpha_0$. The hatched region is the region of bound, pulsating solutions when $\alpha_0 = 0$.}
\end{figure}

Using the above considerations, one can construct the landscape of homologous harmonic-enthalpy solutions shown in Figure~\ref{fig:TypeIIlandscape3}. This figure is a comprehensive overview of the results of this analysis and can be understood as follows. The dashed line in Figure~\ref{fig:TypeIIlandscape3} is the locus of hydrostatic solutions ($a = 1$, $\dot{a} = \ddot{a} = 0$). A  radial Lagrangian perturbation of the harmonic-enthalpy equation of motion ($\Delta a = a - 1 \ll 1$) results in the following dispersion relation:
\be
\omega^2 = \left(\gamma - \frac{4}{3}\right)\omega_d^2,
\ee
which is the canonical result that a pressure-supported equilibrium is dynamically unstable for $\gamma < 4/3$ \citep{CM1933}. The non-linear extension of this linear dynamical instability is represented by the collapse and escape regions to the left of $\gamma = 4/3$ in Figure~\ref{fig:TypeIIlandscape3}. Notice that for a purely radial perturbation, the instability leads to collapse for an inward perturbation and to escape for an outward perturbation. Pressure and gravity are in balance in equilibrium, and which force wins out depends upon the direction of the perturbation. A general angularly-dependent perturbation will result in a separation of material from the unstable equilibrium, with underpressure regions ($\delta < 0$) collapsing and overpressure regions ($\delta > 0$) escaping.

For $\gamma > 4/3$, the solutions are linearly stable, and non-linear instability generally requires a large departure from equilibrium. Collapse requires the complete loss of pressure support ($\delta < -1$), since pressure increases more rapidly than gravity under compression, and escape requires $\dot{a}^2 > 0$ as $a \to \infty$, which translates to  
\be\label{eq:deltaesc}
\delta > \delta_{\rm esc} \equiv \frac{3\gamma - 4}{1 + 3(\gamma-1)\alpha_0}.
\ee
Solutions in between these limits are non-linearly stable even for large $\delta$. When $\alpha_0 \ll 1$, the escape of a fully-ionized gas ($\gamma = 5/3$) requires $\delta > 1$, \emph{i.e.}, an overpressure that is twice the equilibrium pressure.\footnote{The escape threshold can also be understood as the point where the pressure work between the minimum and maximum radius exceeds the work done by gravity and magnetic fields.}

As the importance of radiation pressure increases, however, the non-linear stability threshold for escape decreases to small but finite values, shown by the solid lines (labeled by $\alpha_0$) in Figure~\ref{fig:TypeIIlandscape3}. The reduction of the effective adiabatic index towards $4/3$ moves the system toward the region of linear instability in Figure~\ref{fig:TypeIIlandscape3}, and equation (\ref{eq:deltaesc}) quantifies what level of outward departure from equilibrium is required for escape to occur. Notice that radiation pressure only affects the stability threshold for escape. Although the softening of the equation of state with increasing $\alpha_0$ shortens the time to minimum volume, collapse still requires a complete loss of pressure support ($\delta < -1$).

These results can be understood in terms of an energy analysis as follows. A collapse solution has $E \to -\infty$ with $K \to \infty$, an escape solution has $E \to 0$ with $K = E_0 > 0$, and a pulsating solution is a bound solution with $E_0 < 0$ and a periodic transfer between kinetic and non-kinetic energy. Using the harmonic-enthalpy separation constants and the expression for $\delta$ in Table~\ref{tab:Variable_star}, the total non-kinetic energy at $t = 0$ can be expressed equivalently as 
\be\label{eq:E01}
E_0 = \left(E_{g0} + E_{B0}\right)\frac{\delta_{\rm esc} - \delta}{\delta_{\rm esc} + 1},
\ee
\be\label{eq:E02}
E_0 =  \left(E_{m0} + E_{r0}\right)\frac{\delta - \delta_{\rm esc}}{\delta + 1}.
\ee
For hydrostatic equilibrium solutions with negligible radiation pressure ($\delta = 0$ and $\alpha_0 = 0$), these reduce to
\be\label{eq:CFeq1}
E_0 = \frac{3\gamma - 4}{3(\gamma - 1)}\left(E_{g0} + E_{B0}\right),
\ee
\be\label{eq:CFeq2}
E_0 =  -(3\gamma - 4)\left(E_{m0} + E_{r0}\right).
\ee
\cite{CF1953} use equation (\ref{eq:CFeq1}) to argue that $E_{B0} + E_{g0} < 0$ is required for a stable equilibrium when $\gamma > 4/3$, which in turn implies a limit on the magnitude of the internal magnetic field. It can be seen from equation (\ref{eq:CFeq2}), however, that this logic is flawed, since $E_{m0} + E_{r0} > 0$ by definition, and therefore equilibrium solutions always have $E_0 < 0$ when $3\gamma-4 > 0$. Equation (\ref{eq:CFeq1}) then implies that stable equilibrium solutions must have $E_{B0} + E_{g0} < 0$, independent of the magnetic field strength. Although they are not guaranteed to be stable, equilibrium solutions with strong magnetic fields exist, with the pressure and gravitational forces adjusting to compensate. Equations (\ref{eq:E01}) and (\ref{eq:E02}) are the appropriate generalizations to (\ref{eq:CFeq1}) and (\ref{eq:CFeq2}) for non-linear stability and a non-negligible radiation pressure, and they imply that $-1 < \delta < \delta_{esc}$ is a necessary condition for stability, and that $E_{B0} + E_{g0} < 0$.

\subsection{Time dependence of $\gamma = 4/3$ solutions}\label{sec:temporal43}

The $\gamma = 4/3$ temporal equation can be integrated once to give
\be\label{eq:temporal43}
\frac{\dot{a}^2}{2\omega_d^2} = \delta \left(1 - \frac{1}{a}\right),
\ee
the implicit solution to which is \citep{Low1984a}
\[
\sqrt{2\delta} \omega_d t = \pm \sqrt{a(a-1)} \pm \tanh^{-1}\sqrt{\frac{a-1}{a}}
\]
for $\delta > 0$, and
\[
\sqrt{-2\delta} \omega_d t = \pm \sqrt{a(1-a)} \pm \tan^{-1}\sqrt{\frac{1-a}{a}}
\]
for $\delta < 0$. These correspond to escape and collapse solutions, respectively. Note that the integration constant in equation (\ref{eq:temporal43}) has been set to zero, as it was in equation (\ref{eq:temporalHE}), which is consistent with it reducing to the equilibrium solution when $\delta = 0$.\footnote{\cite{GW1980} consider solutions with $\dot{a}_0 \neq 0$, specifically $K_0 = -E_0$, \emph{i.e.}, an initial velocity equal to the escape velocity. These solutions require $\delta < 0$.}

\subsection{Space dependence of $\gamma = 4/3$ solutions}\label{sec:degenerate}

The spatial profiles here and throughout will be taken to be polytropes with polytropic index $n$, where in general $n \neq 1/(\gamma-1)$:
\be
P_0 = P_{c0} \left(\frac{\rho_0}{\rho_{c0}}\right)^{1+1/n}.
\ee
The parameter space for the spatial profiles when $\gamma = 4/3$ is otherwise unconstrained. Only a couple of examples will be provided here. The assumption of constant plasma $\beta$ requires $\chi = 1/3$. Under these assumptions, the spatial component of the equation of motion can be combined with the Poisson equation to give
\be\label{eq:LEconstantBeta}
\left(1+\frac{1}{3\beta}\right)\frac{1}{\xi^2}\dv{}{\xi}\left(\xi^2 \dv{\theta_0}{\xi}\right) + \theta_0^n + \delta\frac{\omega_d^2}{\omega_g^2} = 0,
\ee
where $\theta_0 = T_0/T_{c0}$,
\be
\xi \equiv \frac{r}{R} = \frac{r_0}{R_0},
\ee
and here
\[
R_0 \equiv \sqrt{\frac{P_{c0}(n+1)}{4\pi G \rho_{c0}^2}}
\]
is the usual Lane-Emden length scale. An analytical example for $n = 1$ is
\[
\theta_0 = \left(1+\delta \frac{\omega_d^2}{\omega_g^2}\right)\frac{\sqrt{1 + (3\beta)^{-1}}}{\xi}\sin\left(\frac{\xi}{\sqrt{1 + (3\beta)^{-1}}}\right) - \delta \frac{\omega_d^2}{\omega_g^2},
\]
with $P_0 = P_{c0}\left(\rho_0/\rho_{c0}\right)^2$ and $P_{B0} = P_0/\beta$. It can be seen that the magnetic field expands the polytrope by a factor of $\sqrt{1 + (3\beta)^{-1}}$, and that the non-equilibrium flow results in an overall offset of the solution. These solutions smoothly transition to Lane-Emden solutions as $\beta \to \infty$ and $\delta \to 0$. 

A more general solution set, valid for $\chi \neq 1/3$, can be obtained by defining the Lorentz force as follows:
\[
\lla\bJ \times \bB\rra_{r0} \equiv \epsilon \rho_0 r_0 \omega_d^2,
\]
where the hydrodynamic limit can be recovered for $\epsilon \to 0$. In this case the modified Lane-Emden equation is
\be\label{eq:LEharmonicME}
\frac{1}{\xi^2}\dv{}{\xi}\left(\xi^2 \dv{\theta_0}{\xi}\right) + \theta_0^n + (\delta-\epsilon)\frac{\omega_d^2}{\omega_g^2} = 0.
\ee
It can be seen here that both the magnetic field and non-equilibrium flow generate an overall offset of Lane-Emden solutions. These solutions may have some relevance for the equilibria obtained by \cite{Braithwaite2008}. Viable solutions are nontrivial to find, however, as the radius at which the magnetic field is zero does not automatically coincide with the radius at which the density and pressure profiles are zero, nor is the magnetic pressure profile guaranteed to be positive everywhere inside the polytrope. Unlike the harmonic enthalpy solutions described next, the parameter space for these solutions must be explored mostly numerically.

\subsection{Time dependence of harmonic-enthalpy solutions}\label{sec:temporalHE}

\begin{deluxetable}{c|c}
\tablecaption{Time evolution of spherical pulsations ($\gamma=5/3$).
\label{tab:sph53}}
\tablehead{
\colhead{Quantity} & \colhead{Expression}
}
\setlength{\extrarowheight}{10pt}
\startdata
Time & $\displaystyle t = \frac{\varphi - \delta \sin \varphi}{\omega_d \left(1-\delta\right)^{3/2}}$\\[10pt]
Scale factor & $\displaystyle a = \frac{1-\delta \cos \varphi}{1-\delta}$\\[10pt]
Velocity & $\displaystyle \dot{a} = \omega_d\frac{\delta\sqrt{1-\delta}\sin\varphi}{1-\delta\cos \varphi}$\\[10pt]
Acceleration & $\displaystyle \ddot{a} = \omega_d^2\frac{\delta\left(1-\delta \right)^2\left(\cos \varphi-\delta\right)}{\left(1-\delta \cos \varphi\right)^3}$\\[10pt]
\enddata
\end{deluxetable}

Although the time dependence of the harmonic enthalpy solutions must in general be solved numerically, an analytical solution can be derived for $\gamma = 5/3$. The energy integral in that case is
\be\label{master_velocity_g53}
\dot{a} = \pm\omega_d\sqrt{\left(1-\frac{1}{a}\right)\left(\delta-1+\frac{\delta+1}{a}\right)},
\ee
the solution to which is 
\[
\omega_d t = \sqrt{\frac{\left(1-a\right)\left(a-{\cal A}\right)}{1-\delta}}
+ \frac{2}{(1-\delta)^{3/2}}\tan^{-1}\sqrt{\frac{1-a}{a-{\cal A}}}
\]
for $0 < \delta < 1$ (for $-1 < \delta < 0$ there is a negative sign in front of the square root) and
\[
\omega_d t = \sqrt{\frac{\left(a-1\right)\left(a-{\cal A}\right)}{\delta-1}} + \frac{2}{(\delta-1)^{3/2}}\tanh^{-1}\sqrt{\frac{a-1}{a-{\cal A}}}
\]
for $\delta > 1$, where the radius ratio ${\cal A}$ is defined in Table~\ref{tab:sph53cv}. By defining the following angle variable:
\[
\varphi \equiv 2\tan^{-1}\left(\sqrt{\frac{1-a}{a-{\cal A}}}\right),
\]
the solution for $-1 < \delta < 1$ can be expressed parametrically as shown in Table~\ref{tab:sph53}. The expressions for $t$ and $a$ in Table~\ref{tab:sph53} represent a \emph{curtate trochoid}, with the $\delta \to -1 $ limit being a common trochoid, or cycloid.\footnote{https://en.wikipedia.org/wiki/Trochoid} The other expressions in Table~\ref{tab:sph53} are readily derived from the definitions of those quantities in terms of $a$.
\begin{figure}
\includegraphics[width=\columnwidth]{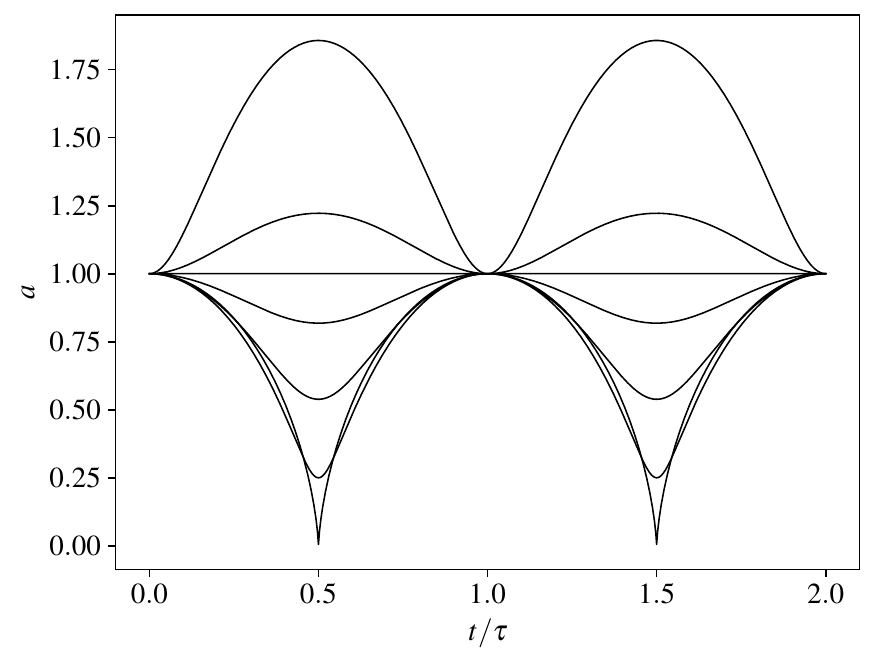}
\caption{\label{fig:Trochoid}Time evolution of spherical pulsating equilibria with $\gamma = 5/3$ and $\delta = -0.99$, $-0.6$, $-0.3$, $-0.1$, $0$, $0.1$, and $0.3$ (from bottom to top).}
\end{figure}
The analytical solution for  spherical pulsating equilibria with $\gamma = 5/3$ is shown in Figure~\ref{fig:Trochoid} for several values of $\delta$. Notice that the further the flow departs from hydrostatic equilibrium, the more it appears to be approaching complete collapse. As noted above, however, collapse will not occur as long as there is non-zero pressure support, because pressure increases more rapidly than gravity for $\gamma > 4/3$.

\begin{deluxetable}{c|c}
\tablecaption{Critical values ($\gamma=5/3$).
\label{tab:sph53cv}}
\tablehead{
\colhead{Quantity} & \colhead{Expression}
}
\setlength{\extrarowheight}{10pt}
\startdata
Radius ratio & $\displaystyle {\cal A} = \frac{1+\delta}{1-\delta}$\\[10pt]
Peak velocity & $\displaystyle v_{\rm max} = R_0\omega_{d}\sqrt{\frac{\delta^2}{1+\delta}}$\\[10pt]
Peak acceleration & $\displaystyle \dot{v}_{\rm max} = R_0\omega_{d}^2\left(-\frac{\delta}{{\cal A}^2},\delta\right), \; \delta\lessgtr 0$\\[10pt]
$\displaystyle v_{\rm max}$ scale factor & $\displaystyle a\left(v_{\rm max}\right) = 1+\delta$\\[10pt]
$\displaystyle \dot{v}_{\rm max}$ scale factor & $\displaystyle a\left(\dot{v}_{\rm max}\right) = \left({\cal A}, 1\right), \; \delta \lessgtr 0 $\\[10pt]
 Pulsation period & $\displaystyle \tau = \frac{2\pi}{\omega_d\left(1-\delta\right)^{3/2}}$\\[10pt]
\enddata
%\begin{tablenotes}
%\item[1] $^\dagger m \in \mathbb{Z}$.
%\end{tablenotes}
\end{deluxetable}

\begin{figure}
\includegraphics[width=\columnwidth]{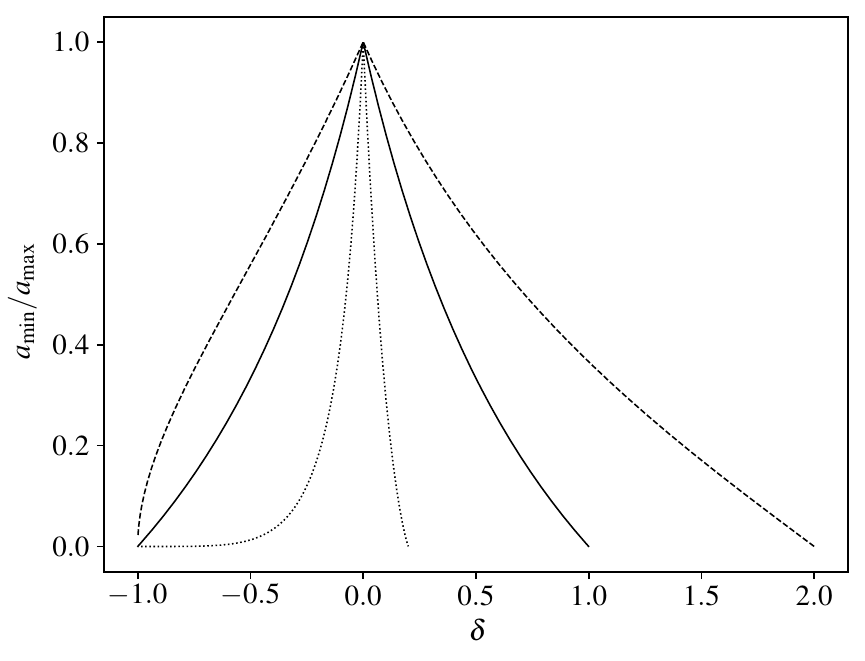}
\caption{\label{fig:a_extrema}Ratio of minimum to maximum radius for spherical pulsating equilibria with $\gamma = 5/3$ (\emph{solid}), $2$ (\emph{dashed}), and $7/5$ (\emph{dotted}).}
\end{figure}

Critical values for this solution are shown in Table~\ref{tab:sph53cv}. 
The radius ratio of a pulsating solution, \textit{i.e.}, the ratio of radial extrema, can be obtained by setting $\dot{a} = 0$ in equation (\ref{master_velocity_g53}); for $\gamma = 5/3$, it can be also be obtained by setting $\varphi = \pi$ in the expression for $a$ in Table~\ref{tab:sph53}. The scale factor at which the velocity and acceleration reach their peak value can be obtained by setting $\ddot{a} = 0$ and $\dddot{a} = 0$, respectively. The pulsation period for $\gamma = 5/3$ is obtained by setting $\varphi = \pi$ in the expression for $t$ in Table~\ref{tab:sph53}. The minimum scale factor as a function of $\delta$ is shown in Figure~\ref{fig:a_extrema} for three values of $\gamma$; the results for $\gamma = 7/5$ were obtained numerically. These curves reflect the fact that a gas with a softer equation of state pulsates more strongly and escapes more easily. Despite approaching the collapse threshold more rapidly, however, a gas with a soft equation of state still requires the complete loss of pressure support ($\delta \leq -1$) to collapse to a singularity.

The $\gamma = 5/3$ solution is directly analogous to a Keplerian orbit. In the absence of pressure and magnetic fields, conservation of angular momentum implies $\Omega = \Omega_0/a^2$, and the equation of motion is
\be\label{eq:keplerian}
\ddot{a} = \frac{\Omega_0^2}{a^3} - \frac{\Omega_\circ^2}{a^2}, \;\; \Omega_\circ^2 \equiv \frac{Gm}{r_0^3},
\ee
which can be seen to be equivalent to the equation of motion in Table~\ref{tab:Variable_star} with $\gamma \to 5/3$, $\omega_d^2 \to \Omega_\circ^2$ and $\delta \to \Omega_0^2/\Omega_\circ^2 - 1$. This is the equation governing Keplerian orbits, with $\delta$ being the orbital eccentricity. The orbital angle is given by
\[
\phi = \int \Omega \, dt = \Omega_0 \int \frac{da}{a^2\dot{a}} =  2\tan^{-1}\left(\sqrt{{\cal A}\frac{1-a}{a-{\cal A}}}\right),
\]
which can be inverted to give
\[
r = \frac{r_0\left(1+\delta\right)}{1 + \delta\cos\phi},
\]
from which it can be seen that $r_0$ (${\cal A} r_0$) is the perigee (apogee) of the orbit, ${\cal A}$ is the ratio of the apsides, and $r_0(1+\delta)$ is the semi-latus rectum. The parameter $\varphi$ in Table~\ref{tab:sph53} is related to $\phi$ by the expression
\[
\tan\frac{\varphi}{2} = \sqrt{\frac{1-\delta}{1+\delta}}\tan\frac{\phi}{2}.
\]
A circular orbit corresponds to $\delta = 0$, $\delta = 1$ corresponds to the orbital speed being equivalent to the escape speed, $\delta = -1$ corresponds to free-fall, and $-1 < \delta < 1$ corresponds to elliptical orbits, with the sign of $\delta$ determining which focus of the ellipse contains the center of gravity.

\subsection{Space dependence of harmonic enthalpy solutions}\label{sec:spatial}

\begin{figure}
\includegraphics[width=\columnwidth]{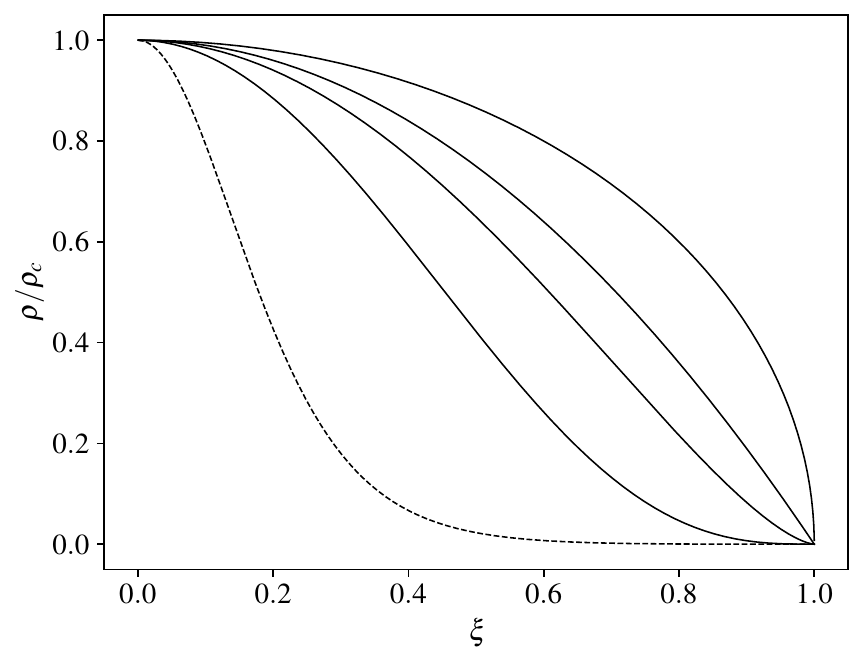}
\caption{\label{fig:density}
Density profiles of magnetic polytropes with $n = 1/2$, $1$, $3/2$, and $3$ (\emph{solid}, from top to bottom) normalized to the central density. The Eddington solution (\emph{dashed}) is shown for comparison.}
\end{figure}

For a harmonic enthalpy profile, the solutions for polytropes have the following simple analytical form:
\[
\theta_0 = 1 - \xi^2, \;\; \rho_0 = \rho_{c0}\left(1 - \xi^2\right)^n,
\]
\[
P_0 = P_{c0}\left(1 - \xi^2\right)^{n+1}, \;\; \Phi_0 = \frac{4\pi G \left(P_{c0} - P_0\right)}{(1+\delta)\omega_d^2} - \frac{Gm}{r_0},
\]
where here
\[
R_0 \equiv \sqrt{\frac{2(n+1)P_{c0}}{(1+\delta)\rho_{c0}\omega_d^2}} = \sqrt{\frac{6}{1+\delta}}\frac{\omega_g}{\omega_d}\sqrt{\frac{(n+1)P_{c0}}{4\pi G\rho_{c0}^2}}.
\]
It follows that the departure from equilibrium for harmonic-enthalpy solutions can be expressed as
\[
\delta = \frac{\Delta P_{0}}{P_{{\rm eq}}} = \frac{P_{0} - P_{{\rm eq}}}{P_{{\rm eq}}}, \;\; P_{\rm eq} = \frac{\rho_{c0}R_0^2\omega_d^2}{2(n+1)}\left(1 - \xi^2\right)^{n+1}.
\]
The enclosed and total polytrope mass are given by
\[
\frac{m}{\rho_{c0} V_0} = \xi^3 \,_2F_1\left(\frac{3}{2}, -n; \frac{5}{2}; \xi^2\right), \;\; V_0 = \frac{4}{3}\pi R_0^3,
\]
\[
M = V_0 \overline{\rho}_0, \;\; \overline{\rho}_0 \equiv \frac{3 \sqrt{\pi}\Gamma(n+1)}{4 \Gamma(n+5/2)}\rho_{c0},
\]
where $\,_2F_1$ is the ordinary hypergeometric function and $\Gamma$ is the gamma function. These results are independent of the form of the magnetic pressure.

\begin{figure}
\includegraphics[width=\columnwidth]{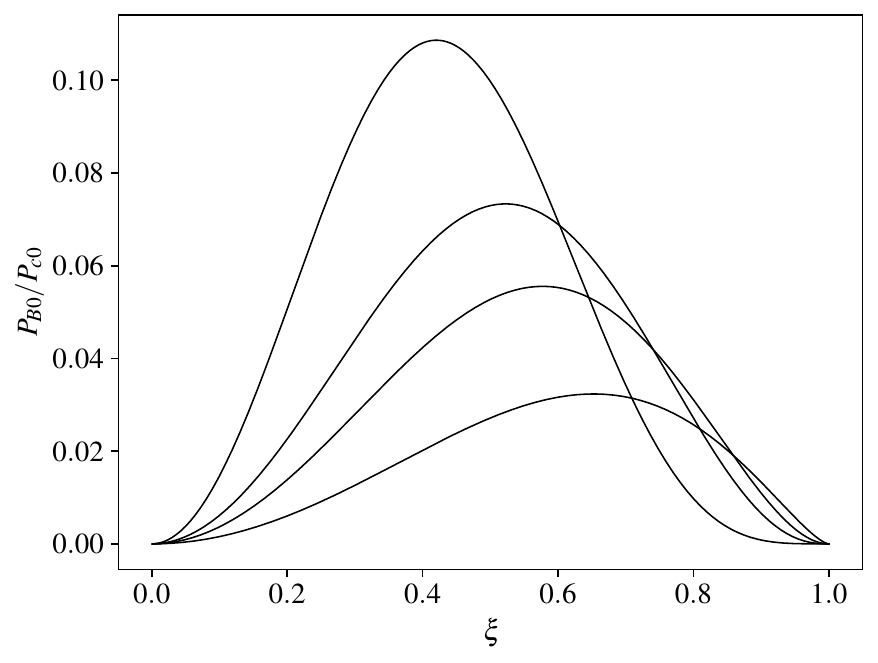}
\caption{\label{fig:PB}Magnetic pressure profiles of harmonic-enthalpy polytropes at $t = 0$ with $\delta = 0$, $\chi = 1$, and $n = 1/2$, $1$, $3/2$, and $3$ (from bottom to top), normalized to the central thermal pressure.}
\end{figure}

The density profile for several values of $n$ is shown in Figure~\ref{fig:density}. An outstanding feature of these magnetic polytropes is that they are much less centrally-condensed than the standard model; the 
central density is ${\sim}3$ times the mean density for  $n = 1$ and ${\sim}7$ times for $n = 3$, as compared to ${\sim}54$ times for the Eddington solution \citep{Chandra1931}. Figure~\ref{fig:PB} shows the corresponding magnetic pressure profiles for $\chi = 1$ and a surface magnetic field of zero. Notice that the solution with the smallest field amplitude ($n = 1/2$) has a density profile that departs the furthest from the Eddington solution, a non-intuitive result that is a reflection of the details of the force balance (see Appendix~\ref{sec:forcebalance}). Despite the fact that the magnetic pressure in these solutions is subthermal, it cannot be regarded as a perturbation on the hydrodynamic force balance. Further details on the magnetic pressure profiles are provided in Appendices~\ref{sec:polytrope} and \ref{sec:forcebalance}.

%\begin{figure}
%\includegraphics[width=\columnwidth]{figs/flux.pdf}
%\caption{\label{fig:flux}Flux-to-mass ratio of harmonic-enthalpy polytropes with $\chi = 1$ and $n = 1/2$, $1$, $3/2$, and $3$ (from bottom to top).}
%\end{figure}

The combined constraints of zero surface field and positive magnetic pressure results in a minimum value for the internal magnetic field for the harmonic-enthalpy solutions given by
\be\label{eq:minB}
\overline{B}_{0} \gtrsim 10^8 \left(\frac{M}{M_\odot}\right) \left(\frac{R_0}{R_\odot}\right)^{-2} \; {\rm G},
\ee
a result that is independent of $\delta$. This expression uses the relation $\overline{B}_{0} = 5\sqrt{P_{B0}}$, valid when the magnetic field is in gauss and the magnetic pressure is in barye. This minimum magnetic field strength corresponds to a mass-to-flux ratio that is comparable to ambient values \citep{McKee2007}. Since this minimum magnetic field is several orders of magnitude larger than observed fields, these solutions can be trivially extended to include a small surface field consistent with observations. Harmonic-enthalpy solutions with a non-negligible surface field exist, but they have surface fields of comparable magnitude to (\ref{eq:minB}) and are more relevant to magnetic fusion. They also invalidate the conclusions discussed here based on the virial theorem, although the virial analysis can be easily extended to include surface terms.\footnote{Surface fields provide a negative contribution to the total energy  \citep{Zweibel1990,MZ1992,FP2000,McLeman2012}.}

The magnetic pressure profiles can be used to relate the escape threshold (\ref{eq:deltaesc}) to the mass of a harmonic-enthalpy polytrope. Although the polytrope solutions are valid for arbitrary $n$ when $\alpha_0 \ll 1$, solutions with non-negligible radiation pressure require constant $\alpha_0$ and therefore $n = 3$. Using the relationship
\[
a_B = \frac{8\pi^5 k_B^4}{15 c^3 h^3},
\]
the following temperature-independent expression can be derived:
\[
\alpha_0\left(1+\alpha_0\right)^3 = \left(\frac{8\pi^5 \mu^4 m_p^4}{45 c^3 h^3}\right)\left(\frac{P_{c0}}{\rho_{c0}^{4/3}}\right)^3,
\]
where 
\[
\frac{P_{c0}}{\rho_{c0}^{4/3}} = (1+\delta)\frac{\pi G}{6}\left(\frac{315 M}{64\pi}\right)^{2/3}\frac{\omega_d^2}{\omega_g^2}
\]
is specific to the $n = 3$ harmonic-enthalpy polytrope. Combining these two expressions and replacing $\alpha_0$ with $\delta_{\rm esc}$ via (\ref{eq:deltaesc}) yields the following quartic for $\delta_{\rm esc}$:
\[
\frac{19845\pi^3}{98304}\left(\frac{\mu^2m_p^2M}{m_P^3}\frac{\omega_d^3}{\omega_g^3}\right)^{2}\left(\frac{\gamma-1}{3\gamma - 4}\delta_{\rm esc}\right)^4 + \frac{\delta_{\rm esc}}{3\gamma - 4} - 1 = 0,
\]
where $m_P = \sqrt{\hbar c/G}$ is the Planck mass. For $\gamma = 5/3$, this is
\[
\frac{245\pi^3}{6144}\left(\frac{\mu^2m_p^2M}{m_P^3}\frac{\omega_d^3}{\omega_g^3}\right)^{2}\delta_{\rm esc}^4 + \delta_{\rm esc} - 1 = 0.
\]

\begin{figure}
\includegraphics[width=\columnwidth]{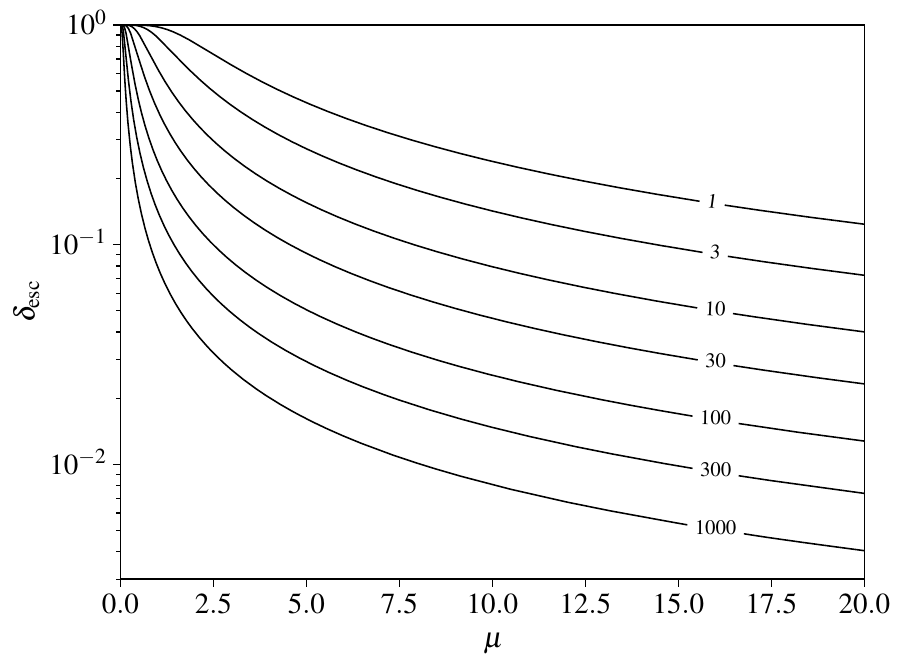}
\caption{\label{fig:Pulsate_delcrit_rad}Escape threshold as a function of mean molecular weight for $\gamma = 5/3$, $n = 3$, and $\chi = 2/3$. The curves are labeled by the polytrope mass in units of $M_\odot$.}
\end{figure}

The value of $\omega_d/\omega_g$ can be related to the surface magnetic field (see Appendix~\ref{sec:polytrope}), and for constant $\chi = 2/3$ and zero surface field is constrained to 
\[
\frac{\omega_{d}^2}{\omega_g^2} = \frac{253}{630}.
\]
Using this value and $m_P^3/m_p^2 \approx 1.8531M_\odot$, the quartic is
\be\label{eq:quartic}
\left(\frac{\mu^2M}{6.54857M_\odot}\right)^2\delta_{\rm esc}^4 + \delta_{\rm esc} - 1 = 0.
\ee
The solutions to equation (\ref{eq:quartic}) are plotted in Figure~\ref{fig:Pulsate_delcrit_rad} for several values of $M$. The reference mass in equation (\ref{eq:quartic}) varies between $5.2660M_\odot$ for $\chi = 1$ ($\omega_d^2/\omega_g^2 = 1024/2205$) and $28.017M_\odot$ for $\chi = 1/2$ ($\omega_d^2/\omega_g^2 = 16/105$). For $\delta_{\rm esc} \ll 1$:
\[
\delta_{\rm esc} \approx \mu^{-1}\left(\frac{m_p^2M}{m_P^3}\right)^{-1/2}\left(\frac{32\sqrt{6
}}{7\pi\sqrt{5\pi}}\right)^{1/2}\left(\frac{\omega_g}{\omega_d}\right)^{3/2},
\]
or
\be\label{eq:delta_approx}
\delta_{\rm esc} \approx 0.15\left(\frac{\mu^2 M}{300M_\odot}\right)^{-1/2}
\ee
for $\chi = 2/3$. The numerical prefactor in equation (\ref{eq:delta_approx}) varies between $0.13$ for $\chi = 1$ and $0.31$ for $\chi = 1/2$.

\subsection{Degenerate matter}

The harmonic-enthalpy solutions motivate an extension to the equation of state for degenerate matter:
\[
P = P_e f(x), \;\; \rho = \rho_e x^3, \;\; x = \frac{p_e}{m_e c},
\]
where 
\[
P_e = \frac{2\pi m_e^4 c^5}{3h^3} \sim 10^{23} \, {\rm barye},
\]
\[
\rho_e  = \frac{8\pi}{3}m_p\mu_e\left(\frac{m_e c}{h}\right)^3 \sim 10^6 \mu_e \, {\rm g \, cm^{-3}},
\]
\[
f(x) = x\left(x^2 - \frac{3}{2}\right)\sqrt{1+x^2} + \frac{3}{2}\sinh^{-1}(x),
\]
$p_e$ is the electron momentum, $\mu_e$ is the average molecular weight per electron, and the other symbols have their usual meanings \citep{Chandra1939}. Notice that $P_e$ and $f$ are a factor of $2$ larger and smaller, respectively, than their usual definitions, so that the pressure has the following form in the non-relativistic and relativistic limits:
\[
\lim_{x\to 0} P = \frac{4}{5}P_e\left(\frac{\rho}{\rho_e}\right)^{5/3}, \;\; \lim_{x\to \infty} P = P_e\left(\frac{\rho}{\rho_e}\right)^{4/3}.
\]
Using $df/dx = 4x^4/\sqrt{1+x^2}$, the spatial pressure equation in Table~\ref{tab:Variable_star} can be solved for the density to give
\be\label{eq:compact}
y = y_{c0} + \left(1-y_{c0}\right)\xi^2, \;\; y \equiv \sqrt{1+\left(\frac{\rho}{\rho_e}\right)^{2/3}},
\ee
where $y_{c0} = y(\rho_{c0})$ and
\[
R_0 = \sqrt{\frac{6P_e\left(y_{c0}-1\right)}{\rho_{c0} \rho_e \pi G}}\frac{\omega_g}{\omega_d}.
\]
The radius has the limits 
\[
\lim_{x\to 0} R_0 = \sqrt{\frac{5P_{c0}}{\rho_{c0} \omega_d^2}}, \;\; \lim_{x\to \infty}  R_0 = \sqrt{\frac{8P_{c0}}{\rho_{c0} \omega_d^2}},
\]
indicating that the solution reduces to the $n = 3/2$ ($n = 3$) harmonic-enthalpy solution in the non-relativistic (relativistic) limit. Since the general solution (\ref{eq:compact}) is not separable, these solutions do not evolve homologously as they transition between the non-relativistic and relativistic limits, and are therefore only valid in the hydrostatic limit or over a narrow range of densities.

\begin{figure}
\includegraphics[width=\columnwidth]{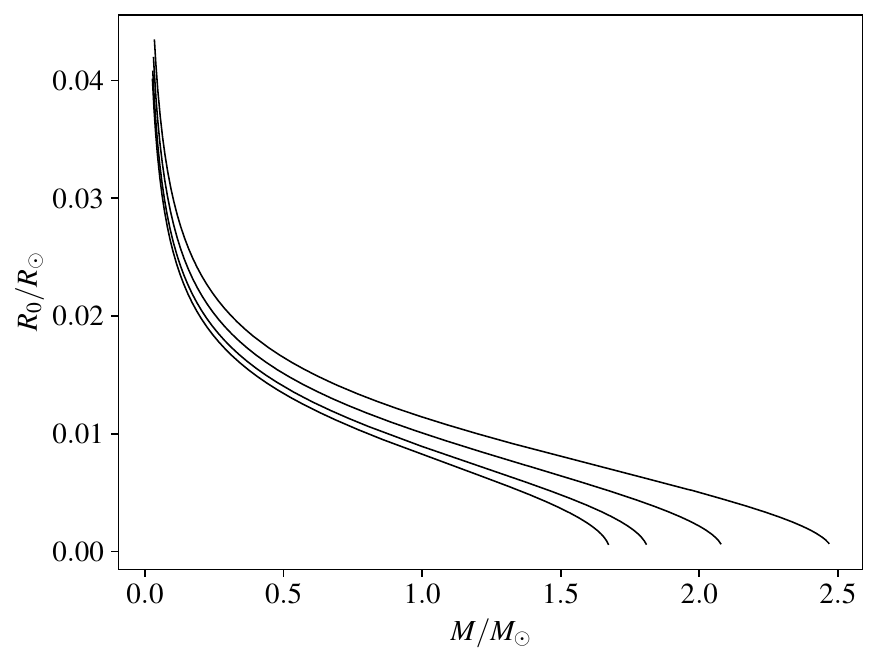}
\caption{\label{fig:WhiteDwarfRadius}Mass-radius relation for harmonic-enthalpy degenerate matter with $\chi = 3/5$, $2/3$, $4/5$, and $1$ (from right to left).}
\end{figure}

\begin{figure}
\includegraphics[width=\columnwidth]{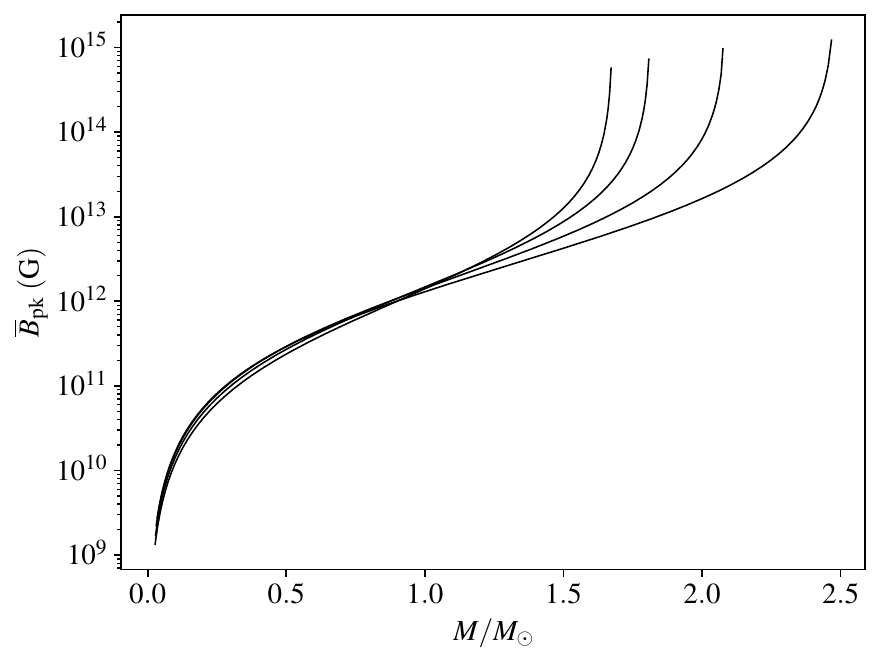}
\caption{\label{fig:WhiteDwarfPeak}Peak internal magnetic field strength for harmonic-enthalpy degenerate matter with $\chi = 3/5$, $2/3$, $4/5$, and $1$ (from right to left).}
\end{figure}

The closed-form solution for the density profile is 
\bea
\frac{\rho}{\rho_e} = \left(\left[\sqrt{1+\left(\frac{\rho_{c0}}{\rho_e}\right)^{2/3}}
\right.\right. \nonumber \\ \left.\left. 
+ \left(1-\sqrt{1+\left(\frac{\rho_{c0}}{\rho_e}\right)^{2/3}}\right)\xi^2\right]^{2} - 1\right)^{3/2},
\eea
and the enclosed mass is given by
\be\label{eq:mWD}
m = 4\pi \rho_{c0} R_0^3 \int \left(\frac{y^2-1}{y_{c0}^2-1}\right)^{3/2} \xi^2 \, d\xi,
\ee
which can be solved analytically in terms of elliptic integrals. For $n = 3$ and $\delta = 0$:
\[
M = \frac{64\pi}{315\rho_e^{2}} \left(\frac{6P_e}{\pi G}\right)^{3/2}\left(\frac{\omega_g}{\omega_d}\right)^{3} = \frac{16\sqrt{6\pi}}{315}\frac{m_P^3}{\mu_e^2 m_p^2}\left(\frac{\omega_g}{\omega_d}\right)^{3}.
\]
Magnetic pressure profiles for constant $\chi$, which look similar to those shown in Figure~\ref{fig:PB}, can be obtained from
\be\label{eq:BWD}
P_{B0} = \frac{8P_e \left(y_{c0}-1\right)}{\xi^{2\nu}\left(2\chi-1\right)} \frac{\rho_{c0}}{\rho_e}\left(-I_1 + \frac{\omega_g^2}{\omega_d^2}I_2\right),
\ee
where $I_1$ and $I_2$ are defined in (\ref{eq:I1I2}). Setting the surface field to zero yields the following constraint:
\[
\frac{\omega_d^2}{\omega_g^2} = \frac{I_{2}(1)}{I_{1}(1)}.
\]
Applying this constraint and solving equations (\ref{eq:mWD}) and (\ref{eq:BWD}) numerically yields the results shown in Figures~\ref{fig:WhiteDwarfRadius} and \ref{fig:WhiteDwarfPeak}. These solutions have peak magnetic fields ${\sim}10^{12}(M/M_\odot)^2$ G and a super-Chandrasekhar maximum mass.

\section{Discussion}\label{sec:discussion}

There are some qualitative similarities between the solutions derived here and the numerical equilibria that have been discovered \citep{BS2004,Braithwaite2008,Braithwaite2009,BS2017,Becerra2022a,Becerra2022b}. For example, the expulsion of magnetic flux from the origin seen in numerical calculations has a natural explanation in terms of satisfying the boundary condition at the origin of a spherical geometry, namely that the magnetic field must be zero at $r = 0$ unless $\chi = 1/3$ to avoid the singularity due to magnetic tension. In addition, Figure 10 of \cite{Braithwaite2008} indicates a slight difference in behavior for solutions with $\chi_r \gtrsim 0.4$ and $\chi_r \lesssim 0.4$, consistent with the analysis in Appendix~\ref{sec:forcebalance}. These similarities, and the minimal assumptions required to derive the solutions, imply that the formalism developed here could be used to guide the search for stable magnetic equilibria in numerical calculations.

The time dependence of the solutions derived here bears a striking resemblance to the breathing modes seen in galaxy formation models; compare Figure~\ref{fig:Trochoid} with Figure 13 of \cite{Hopkins2023a}. Similar solutions may have some relevance to the pulsational behavior seen in the disk solutions of \cite{Riols2017}. Homologous pulsating equilibria exist in cylindrical and planar geometry as well as spherical (a detailed analysis of these is reserved for future work), and in fact exist over a wider region of parameter space because homologous escape solutions are not possible in these geometries. The reason for this is that the gravitational force scales as $a^{-1}$ and $a^0$ in cylindrical and planar geometry, respectively, and therefore the gravitational work always eventually dominates over the work from other forces during expansion.

The primary application of the solutions derived here is as an idealized model for non-linear dynamical stability, and it implies that radial perturbations on the order of $1-10 \%$ have the potential to unbind evolved, high-mass stars that are dominated by radiation pressure. The model requires the assumption of a harmonic-enthalpy spatial profile and an internal mass-to-flux ratio that is comparable to ambient values, such that its direct application to real stars is unclear. At the same time, the derived estimates for non-linear perturbation amplitudes that result in instability may apply more generally. For example, the addition of $1-10 \%$ radial pulsations to the outer envelope of a core collapse supernova progenitor model may improve the robustness of the explosion mechanism \citep{Mezzacappa2026}.

Although addressing the origin of such pulsations is beyond the scope of this analysis, one speculation is that small amplitude pulsations may be natal, \emph{i.e.}, a natural outcome of the star formation process. The formation of high-mass stars is not well understood \citep{McKee2007}, and the original protostellar calculations showed that high-mass protostars exhibit radial pulsations \citep{Larson1969}. While pulsations are damped by both shocks and radiation diffusion, the former are transient and the timescale of the latter can be quite long. For a pulsating equilibrium and bound-free/free-free opacity, which is valid for a wide range of stellar densities and temperatures, the radiation diffusion timescale can be estimated by
\be\label{eq:tradBFFF}
t_{\rm rad} \sim \frac{\rho_0^2 R_0^2}{T_0^{11/2}} a^7.
\ee
This indicates that natal pulsations persist the longest for low-density, large, red stars expanding away from equilibrium ($a > 1$). Delta Cephei, the canonical variable star, has a radius $R_0 \sim 50 R_\odot$ and a velocity curve that indicates a ratio of maximum to minimum radius $a \sim 3$. Assuming solar density and temperature and a solar diffusion timescale of $1.7 \times 10^5$ yr \citep{MS1992}, expression (\ref{eq:tradBFFF}) implies a radiation diffusion time scale for Delta Cephei $t_{\rm rad} \sim 10^{12}$ yr.

Independent of questions associated with the idealized assumptions made here, the equations of motion (\ref{momentumrf})--(\ref{PBeq}) are quite general for a problem with average spherical symmetry, with the only required assumptions being the Cowling approximation and negligible non-radial velocity fluctuations. They thus have much wider applicability than the simplified analysis performed here, and the analysis can be viewed as a demonstration of their utility. Until resolution is no longer a limitation for numerical simulations \citep{Browning2023}, simplified models will continue to provide value, and the rigor of the magnetic model described here is an improvement over existing approaches. In addition, its uncertainty is straightforward to quantify in a full three-dimensional calculation. A measure of the validity of the Cowling approximation is given by
\[
\varepsilon_\Phi \equiv \frac{\lla \rho \pdv{\Phi}{r}\rra}{\lla \rho \rra \pdv{\lla\Phi\rra}{r}} - 1,
\]
and a measure of the importance of non-radial velocity fluctuations is given by
\[
\varepsilon_v \equiv \frac{\lla v^2\rra - \lla v\rra^2}{v_{\rm max}^2},
\]
where $v_{\rm max}$ is the maximum of the adiabatic sound speed, the Alfv\'en speed, and the mean radial velocity.

\newpage

%\section*{Acknowledgments}

%Acknowledge support (or not).

\bibliographystyle{apsrev4-1}

% You should give the same name for your .bbl as your main .tex
% since it is a requirement for posting on ArXiv.
\bibliography{oja_template}

\newpage

\appendix

\section{A. Angular averages}\label{sec:angular_average}

The divergence-free constraint averaged over solid angle is
\[
\frac{1}{r^2}\pdv{}{r}\left(r^2 \frac{1}{4 \pi}\int_0^{2\pi} d\phi \int_0^\pi d\theta \sin \theta \, B_r \right) + \frac{1}{4 \pi r}\int_0^{2\pi} d\phi \int_0^\pi d\theta \, \pdv{}{\theta}\left(B_\theta \sin\theta\right) + \frac{1}{4 \pi r}\int_0^{2\pi} d\phi \int_0^\pi d\theta \, \pdv{B_\phi}{\phi} = 0,
\]
where the middle integral is zero because $\sin\theta = 0$ at $\theta = 0$ and $\pi$, and the third integral is zero because the magnetic field is periodic in $\phi$; $\langle B_\theta \rangle = \langle B_\phi \rangle = 0$ for the same reasons. The divergence-free constraint thus reduces to
\[
\frac{1}{r^2}\pdv{}{r}\left(r^2 \langle B_r\rangle \right) = 0,
\]
which implies $\langle B_r\rangle = 0$ if a finite solution at the origin is required. This result, which is a manifestation of the fact that there are no magnetic monopoles, does not imply $\lla B_r^2\rra = 0$. The fact that total derivatives over $\theta$ and $\phi$ are zero implies that an integration by parts will transfer $\theta$ and $\phi$ derivatives from one factor to another in an integrand and change the sign of the integral. The average radial Lorentz force density is 
\[
\left\langle  \bJ \times \bB \right\rangle_r = - \frac{1}{r^2}\pdv{}{r}\left(r^2   \left\langle P_{B\sphericalangle}\right\rangle\right) + \frac{1}{4\pi r}\int_0^{2\pi} d\phi \int_0^\pi d\theta \, \sin\theta B_\theta\pdv{B_r}{\theta} + \frac{1}{4\pi r}\int_0^{2\pi} d\phi \int_0^\pi d\theta \, B_\phi\pdv{B_r}{\phi},
\]
which is equivalent to
\[
\left\langle  \bJ \times \bB \right\rangle_r = - \frac{1}{r^2}\pdv{}{r}\left(r^2 \left\langle P_{B\sphericalangle}\right\rangle\right) - \frac{1}{4\pi r}\int_0^{2\pi} d\phi \int_0^\pi d\theta \, B_r \pdv{}{\theta}\left(\sin\theta B_\theta\right) - \frac{1}{4\pi r}\int_0^{2\pi} d\phi \int_0^\pi d\theta \, B_r\pdv{B_\phi}{\phi}.
\]
Combining the final two integrals and invoking the divergence-free constraint gives
\[
\left\langle\bJ\times\bB\right\rangle_r = -\frac{1}{r^2}\pdv{}{r}\left(r^2 \left\langle P_{B\sphericalangle}\right\rangle\right) + \frac{1}{4\pi}\int_0^{2\pi} d\phi \int_0^\pi d\theta \, \sin\theta \frac{B_r}{r^2}\pdv{}{r}\left(r^2 B_r\right) = -\frac{1}{r^2}\pdv{}{r}\left(r^2 \left\langle P_{B\sphericalangle}\right\rangle\right) +\frac{1}{r^4}\pdv{}{r}\left(r^4 \left\langle P_{Br}\right\rangle\right).
\]
The same reasoning applied to divergence-free velocity fluctuations yields the following expression for the average radial turbulent force:
\be\label{eq:convection}
-\left\langle \bfv^\prime \cdot \bnabla \bfv^\prime \right\rangle_r = -\left\langle v_r^\prime \pdv{v_r^\prime}{r} + \frac{v_\theta^\prime}{r}\pdv{v_r^\prime}{\theta} + \frac{v_\phi^\prime}{r\sin\theta}\pdv{v_r^\prime}{\phi} - \frac{v_\theta^{\prime 2} + v_\phi^{\prime 2}}{r}\right\rangle = -\frac{1}{r^2}\pdv{}{r}\left(r^2 \sigma_r\right) + \frac{\sigma_\sphericalangle}{r}, \;\; \sigma_i \equiv \left\langle v_i^{\prime 2}\right\rangle,
\ee
a result that along with appropriate closures could be used to extend this formalism to convective equilibria.

\section{B. Virial theorem}\label{sec:virial}

In the absence of surface terms, the magnetic virial theorem (including radiation pressure) is \citep{CF1953}:
\[
\frac{1}{2}\dv{^2I}{t^2} = 2K + 3(\gamma-1)E_m + E_r + E_B + E_g,
\]
where
\[
I = \int r^2 \, dm, \; K = \frac{1}{2}\int v^2 \, dm, \; E_m = \int e_m \, dV, \; E_r = \int e_r \, dV, \; E_B = \int P_B \, dV, \; E_g = - \int \br \cdot \bnabla \Phi \, dm, \; dm = \rho dV.
\]
For a homologous flow, this is
\[
\left(a\ddot{a} + \dot{a}^2\right)\int r_0^2 \rho_0 \, dV_0 =  \dot{a}^2 \int r_0^2 \rho_0 \, dV_0 + \frac{1}{a}\left(\frac{\alpha}{\alpha_0}\right)^{\frac13}\frac{1+\alpha}{1+\alpha_0}\int 3P_0 \, dV_0 + \frac{1}{a}\left(\int P_{B0} \, dV_0 - \int \rho_0 r_0 \frac{d\Phi_0}{dr_0} \, dV_0\right),
\]
or equivalently,
\be\label{virial}
I_0\ddot{a} =\left(\frac{\alpha}{\alpha_0}\right)^{\frac13}\frac{1+\alpha}{1+\alpha_0}\frac{3(\gamma-1)E_{m0} + E_{r0}}{a^2} + \frac{E_{B0} + E_{g0}}{a^2}
\ee
Notice that the moment of inertia and kinetic energy terms both contribute to the virial theorem when there is a time-dependent bulk flow. Comparing this expression to the harmonic-enthalpy equation of motion in Table~\ref{tab:Variable_star} gives expression for $\delta$ in Table~\ref{tab:Variable_star}. Integration of equation (\ref{virial}) gives
\[
K = K_0 + E_0 - E, \;\; {\rm where} \;\; E = E_m + E_r + E_B + E_g.
\]

The virial theorm can be used to provide a constraint on $\omega_d/\omega_g$ due to zero surface field. The various contributions to the total energy for the $n=1$ magnetic polytrope, for example, are
\[
\int_0^{R_0} r_0^2 \rho_0 \, dV_0 = \frac{8\pi}{35} \rho_{c0} R_0^5, \;\;
\int_0^{R_0} 3P_0\, dV_0 = \frac{8\pi}{35} \rho_{c0}R_0^5(1+\delta)\omega_d^2, \;\; 
\int_0^{R_0} \rho_0 r_0 \dv{\Phi_0}{r_0}\, dV_0 = \frac{16\pi}{105}\rho_{c0}R_0^5\omega_g^2,
\]
and
\[
\int_0^{R_0} P_{B0} \, dV_0 = 2\pi \rho_{c0} R_0^5 \left(3-2\nu\right)\left(\frac{2\nu+9}{35[\nu+1][\nu+2]}\omega_d^2 - \frac{2}{105}\frac{2\nu^2 + 15\nu + 34}{[\nu+1][\nu+2][\nu+3]}\omega_{g}^2\right).
\]
Comparison with the equation of motion shows that this requires
\[
\frac{3-2\nu}{(\nu+1)(\nu+2)}\left(\frac{2\nu+9}{4}\omega_d^2 - \frac{1}{6}\frac{2\nu^2 + 15\nu + 34}{\nu+3}\omega_{g}^2\right) - \frac{2}{3}\omega_g^2 = -\omega_d^2 \;\; \to \;\;
\frac{\omega_d^2}{\omega_g^2} = \frac{2(\nu + 6)}{5(\nu+3)}.
\]
As an aside, the equation of motion can also be expressed as a work-energy theorem for a compressible fluid element:
\be\label{specific_work_energy_integral}
\dv{}{t}\left(\frac{v^2}{2} - \frac{W_c}{\rho}\right) = 0, \;\; {W}_c \equiv \rho \int \frac{\bF \cdot \bfv}{\rho} \, dt = \rho \int \frac{\bF \cdot d \br}{\rho}, \;\; \bF \equiv -\bnabla P - \rho \bnabla \Phi + \bJ \times \bB.
\ee

\section{C. Magnetic pressure profiles} \label{sec:polytrope}

The magnetic pressure profile can be derived analytically in most cases, although only a handful of results will be shown here. In order of increasing complexity, these are solutions for 1) $\chi = 1/2$ and general $n$, 2) $n = 1$ and $\chi = 1/3$, 3) $n = 1$ and constant $\chi$, 4) $n = 3$ and $\chi$ varying between $1/3$ and $2/3$, 5) $n = 3$ and constant $\chi$, 6) $n = 1/2$ and $\chi = 1$, and 7) $n = 3/2$ and $\chi = 1$. For constant $\chi$, the average radial Lorentz force is
\be\label{eq:Jr_chi0}
\langle\bJ \times \bB\rangle_r = \frac{2\chi-1}{r^{2\nu}}\pdv{}{r}\left(r^{2\nu} P_{B}\right), \;\;
\nu \equiv \frac{3\chi-1}{2\chi-1}.
\ee
For the harmonic-enthalpy polytropes described in \S\ref{sec:spatial}, using the above expression in the radial force balance gives the following solution for the magnetic pressure profile when $\chi$ is constant:
\be\label{eq:PB_Phi}
\frac{1}{\xi^{2\nu}}\dv{}{\xi} \left(\xi^{2\nu}P_{B0}\right) = \frac{4\pi G\rho_{c0} R_0^2 \rho_0\xi}{3(2\chi-1)}\left(\frac{m}{\rho_{c} V\xi^3}-\frac{\omega_{d}^2}{\omega_{g}^2}\right) \;\; \to \;\; 
\frac{P_{B0}}{P_{c0}} = \frac{2\left(n+1\right)}{(2\chi-1)\left(1+\delta\right)}\left(-I_1 + \frac{\omega_{g}^2}{\omega_{d}^2}I_2\right),
\ee
where
\be\label{eq:I1I2}
I_1 \equiv \frac{1}{\xi^{2\nu}}\int_0^{\xi} \frac{\rho_0}{\rho_{c0}}\xi^{2\nu+1}\, d\xi, \;\; 
I_2 \equiv \frac{1}{\xi^{2\nu}}\int_0^{\xi} \frac{\rho_0}{\rho_{c0}}\xi^{2\nu-2}\frac{m}{\rho_{c} V} \, d\xi.
\ee

1) For $\chi = 1/2$ ($\nu = \pm \infty$), \emph{i.e.}, equal radial and angular magnetic energies, the magnetic pressure gradient term is zero and the equation of motion can be solved algebraically to give
\be
P_{B0} = \rho_0\left(\frac{Gm}{r_0} - r_0^2 \omega_{d}^2\right),
\ee
a solution that has zero surface field and requires 
\[
\omega_{d}^2 \leq \frac{GM}{R_0^3} \;\; \to \;\; \frac{\omega_{d}^2}{\omega_g^2} \leq \frac{\overline{\rho}}{\rho_{c}} = \frac{3 \sqrt{\pi}\Gamma(n+1)}{4 \Gamma(n+5/2)}
\]
to be positive everywhere.

2) For $n=1$ and $\chi = 1/3$, the solution to equation (\ref{eq:PB_Phi}) with zero surface field is
\be
\beta_{c0}^{-1}\left(1-\xi^{2}\left[\frac{5}{2}-2\xi^{2}+\frac{1}{2}\xi^{4}\right]\right)+\frac{\xi^{2}}{2}\left(1-\xi^{2}\right)^{2} \;\;{\rm with} \;\; \frac{\omega_g^2}{\omega_d^2} = \frac{5}{4}\left(1-\beta_{c0}^{-1}\right).
\ee

3) For $n=1$ and constant $\chi$, the solution to equation (\ref{eq:PB_Phi}) with zero surface field is
%\[
%(1+\delta)\frac{P_{B0}}{P_{c0}} = 2\left(3-2\nu\right)\left(\frac{\xi^2}{\nu+1}-\frac{\xi^4}{\nu+2} - \frac{\omega_{g}^2}{\omega_d^2}\left[\frac{\xi^2}{\nu+1} - \frac{8}{5}\frac{\xi^4}{\nu+2} + \frac{3}{5}\frac{\xi^6}{\nu+3}\right]\right).
%\]
\be\label{eq:PBn1}
(1+\delta)\frac{P_{B0}}{P_{c0}} = \frac{3\left(2\nu-3\right)}{(\nu+6)}\xi^2\left(1 - \xi^2\right)^2 = \frac{3\xi^2\left(1 - \xi^2\right)^2}{15\chi - 7} \;\; {\rm with} \;\;\frac{\omega_g^2}{\omega_d^2} = \frac{5(\nu+3)}{2(\nu+6)}.
\ee
Positive $P_{B0}$ requires $\chi > 7/15 \approx 0.467$. In general, solutions with zero surface field can only be obtained for $\chi \gtrsim 2/5$.

4) For $n = 3$ and $\chi = \frac13 \xi^2 + \frac13$, the radial equation of motion is 
\be
\left(2\xi^2-1\right)\dv{P_{B0}}{\xi} + 10\xi P_{B0} = \frac{24 P_{c0}}{1+\delta}\xi\left(1-\xi^2\right)^3\left(\frac{\omega_g^2}{\omega_d^2}\left[1 - \frac{9}{5} \xi^2 + \frac{9}{7} \xi^4 - \frac{1}{3} \xi^6\right]-1\right),
\ee
the solution to which with zero surface field is
\be
(1+\delta)\frac{P_{B0}}{P_{c0}} = \frac{(1-\xi^2)^4\left(1727-1534\xi^2+455\xi^4\right)}{12256} \;\; {\rm with} \;\;  \frac{\omega_g^2}{\omega_d^2} = \frac{23205}{6128}.
\ee
This solution has $\beta_c \approx 0.3$.

5) For $n=3$ and constant $\chi$, the solution to equation (\ref{eq:PB_Phi}) with zero surface field is
%\bea
%(1+\delta)\frac{P_{B0}}{P_{c0}} = 4\left(3-2\nu\right)\left(\frac{\xi^2}{\nu+1} - \frac{3\xi^4}{\nu+2} + \frac{3\xi^6}{\nu+3} - \frac{\xi^8}{\nu+4} \right.
%\nonumber \\ \left. - \frac{\omega_{g}^2}{\omega_d^2}\left[\frac{\xi^{2}}{\nu+1} - \frac{24}{5}\frac{\xi^{4}}{\nu+2} + \frac{339}{35}\frac{\xi^{6}}{\nu+3} - \frac{1112}{105}\frac{\xi^{8}}{\nu+4} + \frac{233}{35}\frac{\xi^{10}}{\nu+5} - \frac{16}{7}\frac{\xi^{12}}{\nu+6} + \frac{1}{3}\frac{\xi^{14}}{\nu+7}\right]\right). \nonumber
%\eea
\be
(1+\delta)\frac{P_{B0}}{P_{c0}} = \frac{(\nu-3/2)\xi^2(1-\xi^2)^4\left(6258+1417\nu+89\nu^2 - 20[\nu+5][5\nu+42]\xi^2 + 35[\nu+5][\nu+6]\xi^4\right)}{2\nu^3+48\nu^2+445\nu+1974},
\ee
with
\be
\frac{\omega_{g}^2}{\omega_d^2} = \frac{105(\nu+5)(\nu+6)(\nu+7)}{8\left(2\nu^3+48\nu^2+445\nu+1974\right)} = \frac{105(13\chi-6) (15\chi-7) (17\chi-8)}{8\left(22050 \chi^3 - 31870 \chi^2 + 15361 \chi - 2469\right)}.
\ee
In this case positive $P_{B0}$ requires $\chi \gtrsim 0.482$ (the real solution to the cubic equation in the denominator of the above expression). 

6) For $n=1/2$ and $\chi = 1$, the solution to equation (\ref{eq:PB_Phi}) with zero surface field is
\bea
(1+\delta)\frac{P_{B0}}{P_{c0}} = \frac{8}{35\xi^4}\left(\left[1-\xi^{2}\right]^{\frac{3}{2}}\left[1 + \frac{3}{2}\xi^2 + \frac{15}{8}\xi^4\right] - 1\right) \nonumber \\ 
 + \frac{9}{128\xi^2}\frac{\omega_g^2}{\omega_d^2}\left(\sqrt{1-\xi^2}\left[1-2\xi^2\right] - \frac{\sin^{-1}\xi}{\xi}\right)^2  \;\; {\rm with} \;\; \frac{\omega_g^2}{\omega_d^2} = \frac{4096}{315\pi^{2}}.
\eea

\begin{figure}
  \centering
    \centering
\includegraphics[width=.48\linewidth]{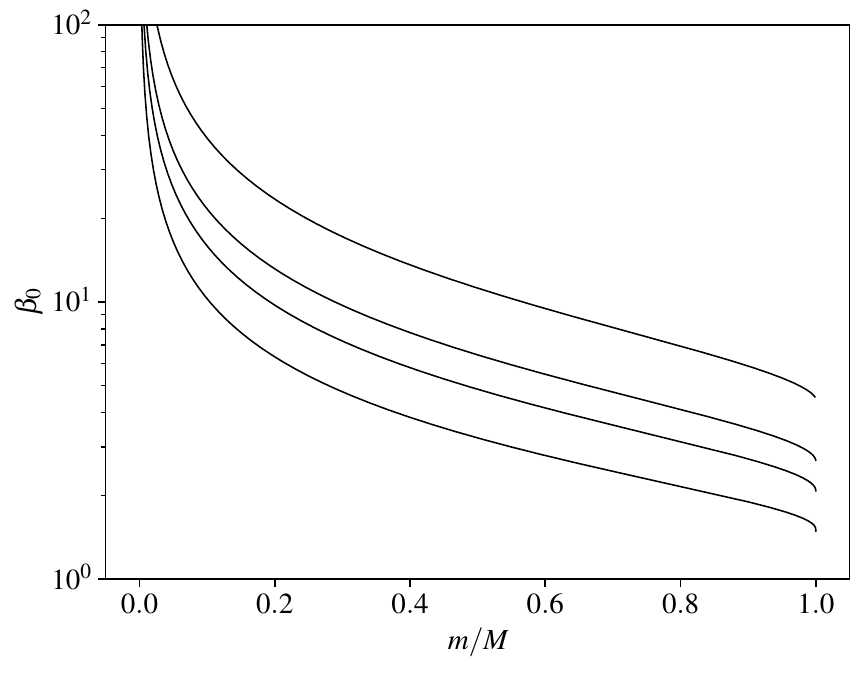}
  \quad
    \centering
\includegraphics[width=.47\linewidth]{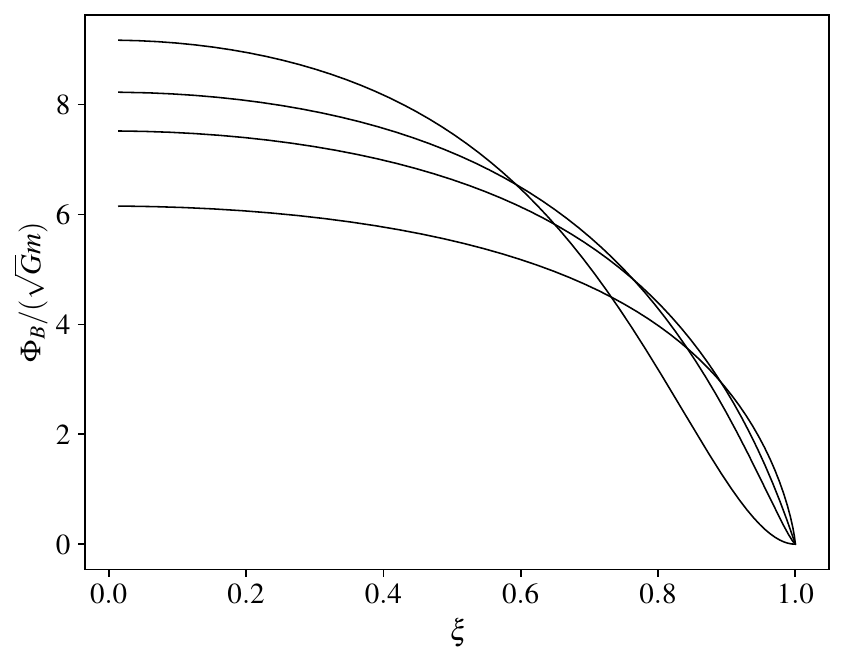}
  \caption{Plasma $\beta$ (\emph{left}) at $t = 0$ (with $\delta = 0$) and flux-to-mass ratio (\emph{right}) of harmonic-enthalpy polytropes, with $\chi = 1$ and $n = 1/2$, $1$, $3/2$, and $3$ (from top to bottom on the left and from bottom to top on the right).}
  \label{fig:beta_flux}
\end{figure}

7) For $n=3/2$ and $\chi = 1$, the solution to equation (\ref{eq:PB_Phi}) with zero surface field is
\bea
(1+\delta)\frac{P_{B0}}{P_{c0}} = \frac{8}{63\xi^4}\left(\left[1-\xi^{2}\right]^{\frac{5}{2}}\left[1 + \frac{5}{2}\xi^2 + \frac{35}{8}\xi^4\right] - 1\right) \nonumber \\ 
 + \frac{32}{63\pi^2\xi^2}\left(\sqrt{1-\xi^2}\left[1-\frac{14}{3}\xi^2+\frac{8}{3}\xi^4\right] - \frac{\sin^{-1}\xi}{\xi}\right)^2  \;\; {\rm with} \;\; \frac{\omega_g^2}{\omega_d^2} = \frac{16384}{945\pi^{2}}.
\eea
Figure~\ref{fig:beta_flux} shows the local plasma $\beta$ at $t = 0$ and the time-independent flux-to-mass ratio for these solutions. It can be seen that $\beta_0$ approaches unity in the outer regions of the star, and that the local mass-to-flux ratio is comparable to ambient values \citep{McKee2007}.

\section{Force Balance}\label{sec:forcebalance}

\begin{figure}
  \centering
    \centering
\includegraphics[width=.47\linewidth]{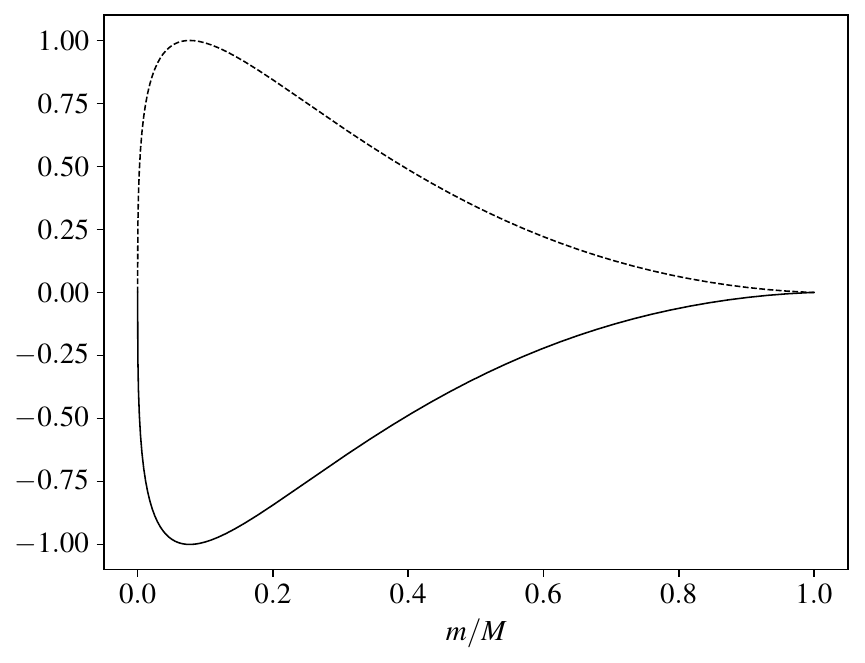}
  \quad
    \centering
\includegraphics[width=.47\linewidth]{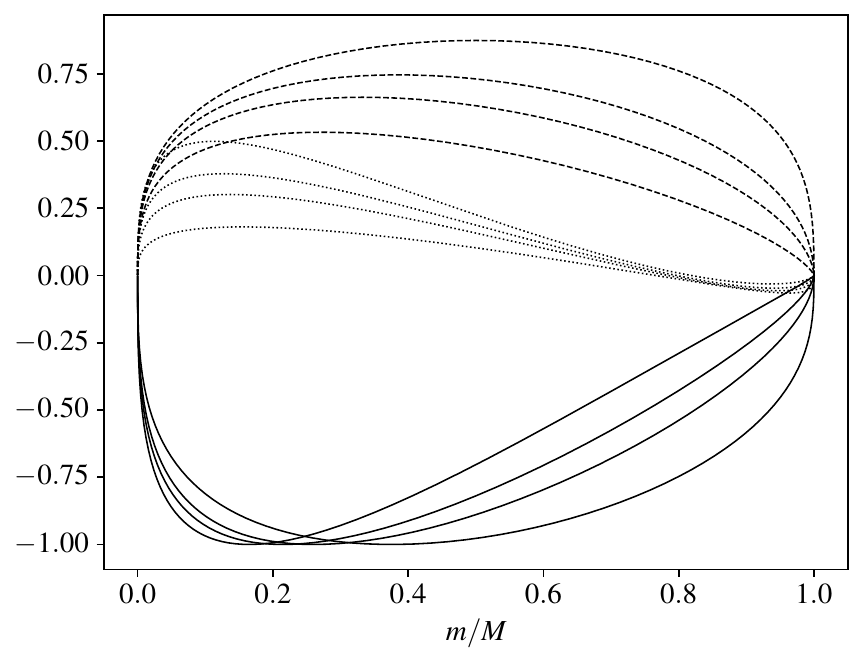}
  \caption{Force balance between gravitational (\emph{solid}) and thermal pressure (\emph{dotted}) for the Eddington solution (\emph{left}), along with the force balance among gravity (\emph{solid}, from bottom to top), thermal pressure (\emph{dashed}, from top to bottom), and magnetic fields (\emph{dotted}, from bottom to top) for the harmonic-enthalpy  polytropes (\emph{right}) with $n = 1/2$, $1$, $3/2$, and $3$.}
  \label{fig:Forces}
\end{figure}

For the harmonic-enthalpy force balance,
\be
F_{B0} = \rho_0\frac{Gm}{r_0^2} - \rho_0 r_0\omega_d^2 = \rho_0\frac{Gm}{r_0^2} + \frac{1}{1+\delta}\dv{P_0}{r_0}.
\ee
The radial Lorentz force can be positive or negative depending on the relative magnitude of the gravitational and pressure/inertial forces, which have the opposite sign. For the solutions described here, $F_{B0} > 0$ everywhere requires $\omega_d^2 < GM/R_0^3$. For values of $\omega_d^2$ larger than this, the Lorentz force in the outer regions of these solutions aids rather than impedes gravity, \emph{i.e.}, the magnetic field contributes to gravitational confinement. The force balance for several values of $n$ are shown in Figure~\ref{fig:Forces}, where it can be seen that the Lorentz force is slightly negative in the outer regions of the equilibria. It can also be seen from this figure that the force balance differs qualitatively from the force balance in the Eddington solution.

%\begin{figure}
%\includegraphics[width=\columnwidth]{figs/beta.pdf}
%\caption{\label{fig:beta}Plasma $\beta$ of harmonic-enthalpy magnetic polytropes at $t = 0$ with $\delta = 0$, $\chi = 1$, and $n = 1/2$, $1$, $3/2$, and $3$ (from top to bottom).}
%\end{figure}

A magnetic pressure minimum near the origin implies an additional constraint on $\omega_d^2$. Assuming that there is not a cusp at the origin (which is equivalent to the assumption that the enclosed mass goes to zero at the origin), it follows that the form of the magnetic pressure near the origin is quadratic:
\[
\lim_{r_0\to 0} P_{B0} = c_B r_0^2 \;\; {\rm with} \;\; c_B > 0.
\]
Using expression (\ref{eq:Jr_chi0}), the radial derivative of the radial Lorentz force density near the origin can be shown to be 
\[
\lim_{r_0\to 0} \dv{F_{B0}}{r_0} = 2\left(5\chi-2\right)c_B,
\]
which for $c_B > 0$ implies
\be\label{eq:chi25}
\lim_{r_0\to 0} \dv{F_{B0}}{r_0} \lessgtr 0 \;\; {\rm for} \;\; \chi \lessgtr \frac{2}{5}.
\ee
This result applies quite generally, even when $\chi$ is not a constant, since by the same arguments both $P_{B\sphericalangle}$ and $P_{Br}$ must be quadratic near the origin with a ratio given by $\chi = \chi(0)$ (for $\chi \neq 1/3$). For the harmonic-enthalpy force balance,
\[
\dv{F_{B0}}{r_0} = \dv{\rho_0}{r_0}\left(\frac{Gm}{r_0^2} - r_0 \omega_d^2\right) + \rho_0\left(\frac{G}{r_0^2}\dv{m}{r_0} - \frac{2Gm}{r_0^3} - \omega_d^2\right),
\]
where the enclosed mass near the origin scales as
\[
\lim_{r_0 \to 0} m(r_0) = \lim_{r_0 \to 0} 4\pi \int \rho_0 r_0^2 \, dr_0 = \frac{4}{3} \pi \rho_{c0} r_0^3.
\]
This along with $d\rho_0/dr_0 = 0$ at $r_0 = 0$ implies
\[
\lim_{r_0\to 0} \dv{F_{B0}}{r_0} = 4\pi G \rho_{c0}^2 - 2\frac{4\pi G \rho_{c0}^2}{3} - \rho_{c0}\omega_d^2 = \rho_{c0}\left(\omega_g^2 - \omega_d^2\right),
\]
which in turn implies the following constraints on the dynamical frequency:
\be\label{eq:constraint1}
0 \leq \omega_d^2 \leq \omega_g^2 \;\; {\rm for} \;\; \chi \geq \frac{2}{5},
\ee
and
\be\label{eq:constraint2}
\omega_g^2 \leq \omega_d^2 \leq \infty \;\; {\rm for} \;\; \chi \leq \frac{2}{5},
\ee
Notice that the dynamical time scale can in principle take on any value, and that it is longer (shorter) than the free-fall time scale when the Lorentz force near the origin is directed outwards (inwards). This is a reflection of the fact that the magnetic field aids (impedes) gravity when $\chi < 2/5$ ($\chi > 2/5$). In the limit $\omega_{d}^2 \gg \omega_g^2$, the Lorentz force dominates over the gravitational force and the polytropes are confined by magnetic fields alone. The solutions in this limit may have some relevance for magnetic fusion but will not be discussed further here.


%\vspace{0.5in}

%\bibliography{ms}{}
%\bibliographystyle{aasjournal}



\end{document}